\documentclass[aip, amssymb, reprint, showpacs]{revtex4-1}

\input{glyphtounicode}
\usepackage{array}
\usepackage{graphicx}
\usepackage{dcolumn}
\usepackage{bm}
\usepackage{float}
\usepackage{amsmath,amsfonts}
\usepackage{url}
\usepackage{setspace}
\usepackage{lineno}
\usepackage{afterpage}

\usepackage{layouts}

\usepackage[nopar]{lipsum}

\usepackage[percent]{overpic}

\usepackage[hidelinks]{hyperref}
\usepackage{amsmath}
\usepackage{amssymb}
\usepackage{amsthm} 
\usepackage{tikz}
\usepackage{color}
\usepackage{graphics} 
\usepackage{multirow} 
\usepackage{graphicx} 
\usepackage[english]{babel}

\usepackage{wrapfig}

\usepackage{cleveref}
\crefformat{figure}{Fig.~#2#1#3}
\Crefformat{figure}{Fig.~#2#1#3}
\crefmultiformat{figure}{Figs.~#2#1#3}{ and~#2#1#3}{, #2#1#3}{ and~#2#1#3}
\Crefmultiformat{figure}{Figs.~#2#1#3}{ and~#2#1#3}{, #2#1#3}{ and~#2#1#3}
\crefrangeformat{figure}{Figs.~#3#1#4 to~#5#2#6}
\Crefrangeformat{figure}{Figs.~#3#1#4 to~#5#2#6}

\crefformat{equation}{Eq.~#2(#1)#3}
\Crefformat{equation}{Eq.~#2(#1)#3}
\crefmultiformat{equation}{Eqs.~#2(#1)#3}{ and~#2(#1)#3}{, #2(#1)#3}{ and~#2(#1)#3}
\Crefmultiformat{equation}{Eqs.~#2(#1)#3}{ and~#2(#1)#3}{, #2(#1)#3}{ and~#2(#1)#3}
\crefrangeformat{equation}{Eqs.~#3#1#4 to~#5#2#6}
\Crefrangeformat{equation}{Eqs.~#3#1#4 to~#5#2#6}

\crefformat{section}{section~#2#1#3}
\Crefformat{section}{Section~#2#1#3}
\crefmultiformat{section}{\S#2#1#3}{ and \S#2#1#3}{, \S#2#1#3}{, and \S#2#1#3}
\Crefmultiformat{section}{\S#2#1#3}{ and \S#2#1#3}{, \S#2#1#3}{, and \S#2#1#3}

\usepackage{cancel}

\usepackage{subfloat}

\usepackage{array}
\usepackage{setspace}

\newlength{\colwidth}
\newlength{\hbarunit}
\newlength{\firstcolwidth}
\newlength{\rowwidth}
\newcolumntype{C}{>{\centering\arraybackslash} m{\colwidth}}

\usepackage[absolute]{textpos}
\usepackage[margin,draft,mode=multiuser]	{fixme}

\fxusetheme{colorsig}

\FXRegisterAuthor{carlo}{acarlo}{Carlo}
\FXRegisterAuthor{jeff}{ajeff}{Jeff}
\FXRegisterAuthor{note}{anote}{Note}
\FXRegisterAuthor{done}{adone}{Done}
\FXRegisterAuthor{read}{aread}{Read}

\newcommand\Pran{\mbox{\it Pr}}

\newcommand\Imag{\mbox{\it Im}}
\newcommand{\charlesx}{CharLES$^X$}

\newcommand\secondcoefficient{\mu'}

\newcommand\mubvib{\mu_{B, \textrm{vib}}}
\newcommand\mubrot{\mu_{B, \textrm{rot}}}
\newcommand\alpharot{\alpha_{\textrm{rot}}}
\newcommand\alphacl{\alpha_{\textrm{sk}}}
\newcommand\Tvibk{T_{\textrm{vib}, k}}

\newcommand\Otwo{\textrm{O}_2}
\newcommand\Ntwo{\textrm{N}_2}

\newcommand\Rgas{R_{gas}}

\newcommand{\pressurelist}{pressures $p_0=$ 0.1 atm \legendshortdashed{}, 1.0 atm \legenddashed{}, 10.0 atm \legendlongdashed{}, and 100.0 atm \legendline{}}

\newcommand\onecolumnwidth{\linewidth} 

\newcommand\twocolumnwidth{0.85\linewidth} 
\usepackage{tikz}
\usetikzlibrary{shapes}
\usetikzlibrary{plotmarks}

\definecolor{darkgray}{rgb}{0.5, 0.5, 0.5}
\definecolor{lightgray}{rgb}{0.8, 0.8, 0.8}

\newcommand{\legendline}[1][black]{\mbox{(\tikz[baseline=-0.75ex,color=#1]{\draw[very thick,fill=#1]  (0,0) -- (2ex,0);})}}
\newcommand{\legendshortdashed}[1][black]{\mbox{(\hspace{-0.02cm}\tikz[baseline=-0.75ex, color=#1, dash pattern=on 1.5pt off 6pt, very thick]{\draw  (0,0) -- (3ex,0);}\hspace{-0.03cm})}}
\newcommand{\legenddashed}[1][black]{\mbox{(\hspace{-0.02cm}\tikz[baseline=-0.75ex, color=#1, dashed, very thick]{\draw  (0,0) -- (4ex,0);}\hspace{-0.03cm})}}
\newcommand{\legendlongdashed}[1][black]{\mbox{(\hspace{-0.02cm}\tikz[baseline=-0.75ex, color=#1, dash pattern=on 4pt off 1pt, very thick]{\draw  (0,0) -- (5ex,0);}\hspace{-0.03cm})}}
\newcommand{\legenddasheddotted}[1][black]{\mbox{(\hspace{-0.02cm}\tikz[baseline=-0.75ex, color=#1, dashed, very thick]{\draw  (0,0) -- (1.4ex,0);}\hspace{-0.03cm}\tikz[baseline=-0.75ex,color=#1]{\draw[very thick,fill=#1]  (0,0) -- (0.25ex,0);}\tikz[baseline=-0.75ex, color=#1, dashed, very thick]{\draw  (0,0) -- (1.4ex,0);}\hspace{-0.03cm}\tikz[baseline=-0.75ex,color=#1]{\draw[very thick,fill=#1]  (0,0) -- (0.25ex,0);}\hspace{-0.00cm})}}

\newcommand{\legendverts}[1][black]{\mbox{(\hspace{-0.02cm}\tikz[baseline=-0.75ex,color=#1]{\draw[very thick,fill=#1]  (0,0) -- (0.15ex,0);}\hspace{-0.01cm}\tikz[baseline=-0.75ex,color=#1]{\draw[very thick,fill=#1]  (0,0) -- (0.15ex,0);}\hspace{-0.01cm}\tikz[baseline=-0.75ex,color=#1]{\draw[very thick,fill=#1]  (0,0) -- (0.15ex,0);}\hspace{-0.01cm}\tikz[baseline=-0.75ex,color=#1]{\draw[very thick,fill=#1]  (0,0) -- (0.15ex,0);}\hspace{-0.01cm}\tikz[baseline=-0.75ex,color=#1]{\draw[very thick,fill=#1]  (0,0) -- (0.15ex,0);}\hspace{-0.01cm}\tikz[baseline=-0.75ex,color=#1]{\draw[very thick,fill=#1]  (0,0) -- (0.15ex,0);}\hspace{-0.00cm})}}

\newcommand{\legenddot}[1][white]{\mbox{(\tikz[baseline=-0.5ex]{\draw[fill=#1](0.25ex,0.25ex) circle (0.75ex); })}}

\newcommand{\legenddotSmultiple}[1][white]{\mbox{(\tikz[baseline=-0.5ex]{\draw[fill=#1](0.25ex,0.25ex) circle (0.25ex); } \tikz[baseline=-0.5ex]{\draw[fill=#1](0.25ex,0.25ex) circle (0.25ex); } \tikz[baseline=-0.5ex]{\draw[fill=#1](0.25ex,0.25ex) circle (0.25ex); })}}
\newcommand{\legendsquare}[1][white]{\mbox{(\tikz[baseline=-0.5ex]{\node[rotate=180,fill=blue,color=#1,fill=black,draw=white,stroke=yellow] at (0.0ex,0.25ex){\pgfuseplotmark{square*}};}\hspace{0.cm})}}

\newcommand{\legendtriangles}[1][black]{(\tikz[baseline=-0.4ex]{\hspace{-0.00cm}\node[rotate=180] at (0.0ex,0.25ex) {\pgfuseplotmark{triangle*}};}\hspace{-0.00cm})}

\newcommand{\legendx}[1][black]{($\times$)}

\newcommand{\tikztriangle}[1][black]{\raisebox{0pt}{\tikz{\node[draw,minimum
			width=0.3cm,minimum height=0.3cm,inner sep=0pt,regular polygon, regular polygon sides=3,fill=#1,rotate=180,thick](){};}}}

\newcommand{\tikzsquare}[1][black]{\raisebox{0pt}{\tikz{\node[draw,minimum
			width=0.3cm,minimum height=0.3cm,inner sep=0pt,regular polygon, regular polygon sides=4,fill=#1,rotate=180,thick](){};}}}

\newcommand{\tikzpentagon}[1][black]{\raisebox{0pt}{\tikz{\node[draw,minimum
			width=0.3cm,minimum height=0.3cm,inner sep=0pt,regular polygon, regular polygon sides=5,fill=#1,rotate=180,thick](){};}}}

\graphicspath{ {./test/}{./figures/} }

\begin{document}

\preprint{JASA/123}
		
\title{Bulk viscosity model for near-equilibrium acoustic wave attenuation} 

\author{Jeffrey Lin}			
\email{linjef@stanford.edu}
\thanks{Corresponding author.}
\affiliation{Department of Electrical Engineering,  
	Stanford University, Stanford, CA, 94305 USA}
\author{Carlo Scalo}
\affiliation{
	\mbox{School of Mechanical Engineering,
	Purdue University, West Lafayette, IN, 47907 USA}
}
\author{Lambertus Hesselink}
\affiliation{Department of Electrical Engineering,  
	Stanford University, Stanford, CA, 94305 USA}

\date{\today} 
		
\begin{abstract}
Acoustic wave attenuation due to vibrational and rotational molecular relaxation, under simplifying assumptions of near-thermodynamic equilibrium and absence of molecular dissociations, can be accounted for by specifying a bulk viscosity coefficient $\mu_B$. 
In this paper, we propose a simple frequency-dependent bulk viscosity model 
which, under such assumptions, accurately captures wave attenuation rates from infrasonic to ultrasonic frequencies in Navier--Stokes and lattice Boltzmann simulations. 
The proposed model can be extended to any gas mixture for which molecular relaxation timescales and attenuation measurements are available. 
The performance of the model is assessed for air by varying the base temperature, pressure, relative humidity $h_r$, and acoustic frequency. 
Since the vibrational relaxation timescales of oxygen and nitrogen are a function of humidity, 
for certain frequencies an intermediate value of $h_r$ can be found which maximizes $\mu_B$. 
The contribution to bulk viscosity due to rotational relaxation is verified to be a function of temperature, confirming recent findings in the literature. 
While $\mu_B$ decreases with higher frequencies, its effects on wave attenuation become more significant, as shown via a dimensionless analysis. 
The proposed bulk viscosity model is designed for frequency-domain linear acoustic formulations but is also extensible to time-domain simulations of narrow-band frequency content flows. 
\end{abstract}

\maketitle
	

\section{\label{sec:intro} Introduction}

The viscous stress tensor of an isotropic Newtonian fluid reads
\begin{align}
\label{e:stresstensor}
\sigma_{ij} = -p \delta_{ij} + \tau_{ij} = \left(-p + \secondcoefficient  \frac{\partial u_k}{\partial x_k}\right) \delta_{ij} + 2\mu S_{ij} \, ,
\end{align}
where $p$ is the thermodynamic pressure, $u_i$ and $x_i$ are the velocity component and spatial coordinate along the $i$-th direction, 
$S_{ij} = 1/2 \left(\partial u_i / \partial x_j+\partial u_j / \partial x_i\right)$ is the strain-rate tensor, 
and $\secondcoefficient$ is the second coefficient of viscosity
\begin{align}
\secondcoefficient  = \mu_B - \frac{2}{3} \mu \, ,
\label{e:secondviscosity}
\end{align}
where $\mu_B$ and $\mu$ are the bulk (or volume) and shear (or dynamic) viscosity, respectively. 
While shear viscosity is a well-characterized fluid property, bulk viscosity is neglected in most fluid problems due to its often unknown value. 
The commonly adopted Stokes' hypothesis, which holds for dilute, monatomic gases, in fact reads $\secondcoefficient = -\frac{2}{3} \mu$, implying $\mu_B=0$. 
However, for many fluids and flows of interest, $\mu_B$ cannot be neglected.\cite{GravesA_1999}
Furthermore, in the context of wave attenuation modeling, 
bulk viscosity is not interpreted as a pure fluid property, as it depends on flow conditions, e.g., acoustic frequency. 

Experimental and semi-empirical characterizations of wave attenuation in air confirm that bulk viscosity effects are due to molecular relaxation, a non-equilibrium thermodynamic process which is sufficiently important so as to affect acoustic energy but not so intense as to require a full description of molecular-level energy exchange dynamics.\cite{CEnerAJ_JournalAcousticalSocietyAmerica_1952,EvansBS_1972,Pierce_1989,BassBE_1972,BassSZBH_1995,BassSZBH_1996} 
Bulk viscosity can significantly affect the attenuation rate of freely propagating waves in mixtures composed of multi-atomic species---for example,  $\mu_B \simeq 2000 \mu$ for $\textrm{CO}_2$\cite{Cramer_2012,GravesA_1999} or for dry air at 110 Hz---
and its effect accumulates over each cycle of acoustic wave propagation, making it very important at higher (e.g., ultrasonic) acoustic frequencies. 
The accurate knowledge of bulk viscosity values is desired in high-fidelity simulations of aeroacoustics problems, with applications including 
thermoacoustic energy conversion and imaging, acoustic energy transfer (AET)\cite{RoesHD_2011}, ultrasonic air-coupled non-destructive examination, and supersonic flow. \cite{ChikitkinRTU_AppliedNumericalMathematics_2015}
In other applications, such as for acoustic trapping, the accurate prediction of heat generation due to both shear and bulk viscosity specifically is critical.\cite{Nyborg_1986}
Bulk viscosity effects are also very important in the presence of strong spatial density gradients, such as in shock waves, which yield high (negative) values of the velocity divergence term in \cref{e:stresstensor}. 
The distinction between losses due to shear or bulk viscosity effects is also relevant to acoustic problems in which the phasing of dilatation can be decoupled from that of pressure. 

Numerical simulations of wave attenuation due to bulk viscosity effects and estimates of bulk viscosity coefficient values have been attempted at various scales of fidelity by previous authors. 
Eu and Ohr\cite{EuO_2001} have shown that Navier--Stokes simulations without bulk viscosity fail to fully capture absorption and dispersion for polyatomic gases, and Claycomb et al.\cite{ClaycombGCCB_2008,ClaycombG_2008} have shown that under rarefied flow conditions and Mach numbers up to 15, the introduction of bulk viscosity can improve predictions of shock thickness and separation in a double cone geometry.
Wagnild et al.\cite{WagnildCSJ_2012} evaluated vibrational relaxation effects in hypersonic boundary layers using direct numerical simulations, and Valentini et al.\cite{ValentiniSBNC_PhysicsFluids_2015} used direct molecular simulation to predict rotational and vibrational relaxation effects in high-temperature nitrogen.
Salomons et al.\cite{SalomonsLZ_PLoSOne_2016} performed lattice Boltzmann method (LBM) simulations without accounting for bulk viscosity effects, finding that the accuracy of simulated acoustic wave attenuation is limited by LBM stability and dissipation to short propagation distances, low frequencies, and high viscosity.
Viggen\cite{Viggen_2014} developed a LBM approach which captures absorption and dispersion due to both shear and bulk viscosity for a hypothetical single-species fluid, 
with results in agreement with theoretical predictions.
Ern and Giovangigli\cite{ErnG_EurJMechBFluids_1995,ErnDV_2004,ErnG_JournalComputationalPhysics_1995} developed multicomponent transport algorithms to accurately calculate various transport coefficients, including bulk viscosity associated with rotational molecular relaxation. 

To the authors' knowledge, established methods providing the value of $\mu_B$ for the prediction of acoustic wave attenuation rates are currently missing in the literature. 
In this manuscript, we present a model for $\mu_B$ valid in near-equilibrium acoustic energy absorption conditions, which can be adopted in frequency-domain formulations as well as narrow-band time-domain simulations. 

\subsection{\label{sec:theory} Bulk viscosity origins and measurements}

While bulk viscosity is negligible for dilute monatomic gases, as confirmed theoretically and experimentally,\cite{GravesA_1999} it is generally non-zero for polyatomic gases. 
For example, a diatomic molecule has six degrees of freedom: three translational, two rotational, and one vibrational. 
The translational degrees of freedom relax to equilibrium conditions relatively quickly, while 
the rotational and vibrational degrees of freedom have longer relaxation times, hence delaying thermodynamic equilibrium when excited. 
At thermodynamic equilibrium, 
the average energy for each translational and rotational degree of freedom is $k_BT/2$, where $k_B$ is Boltzmann's constant and the fluid temperature $T$ is equal to the apparent translational temperature $T_{\textrm{tr}}$ and rotational temperature $T_{\textrm{rot}}$. 
The corresponding translational and rotational energies are given by
\begin{align}
E_{\textrm{tr}} &= \frac{3}{2} \Rgas T_{\textrm{tr}} \label{e:tre} \\ 
E_{\textrm{rot}} &= \frac{1}{2} \frac{5-3\gamma}{\gamma-1} \Rgas T_{\textrm{rot}} \, ,
\label{e:rote}
\end{align}
where $\Rgas$ is the specific gas constant and $\gamma$ is the ratio of specific heats. 
The vibrational energy associated with the $k$-th molecular species is derived from the Boltzmann distribution and is given by
\begin{align}
E_{\textrm{vib}, k} &= \frac{n_k}{n} \Rgas \Tvibk^* \exp\left(-\frac{\Tvibk^*}{\Tvibk}\right)\, , 
\label{e:vibe}
\end{align}
where $n_k/n$ is the species mole fraction, and $\Tvibk^*$ and $\Tvibk$ 
are, respectively, the characteristic molecular vibration temperature and species vibrational temperature. 
At thermodynamic equilibrium, the fluid temperature $T$ is also equal to the apparent vibrational temperatures $\Tvibk$, and
there is zero net energy exchange among the degrees of freedom. 

However, under thermodynamic non-equilibrium conditions, both vibrational and rotational energies diverge from their equilibrium values, and the process of relaxation to equilibrium conditions is represented by the bulk viscosity coefficient. 
The equivalent relaxation process for translational degrees of freedom is embedded within the shear viscosity coefficient. 
Such attenuative processes are of particular interest in sound propagation, where pressure and velocity fluctuations (i.e. fluctuations in translational molecular energy) establish a cycle of compressions and expansions accompanied by temperature fluctuations. 
However, unlike shear viscosity, absorption of sound is the primary effect of bulk viscosity, and the experimental characterization of bulk viscosity in various materials has predominantly been completed through various forms of acoustic wave attenuation measurements.\cite{Medwin_1954,Greenspan_1959,Madigosky_1967,DukhinG_JChemPhys_2009}

Bulk viscosity, $\mu_B$, as defined in the present paper by  \cref{e:stresstensor,e:secondviscosity}, is sometimes referred to in the literature as volume viscosity\cite{LitovitzD_1965}, dilatational viscosity\cite{ChangUD_1964}, expansion viscosity\cite{Batchelor_1973,Sonin_2001}, 
 or the second coefficient of viscosity \cite{LandauL_2013}.
The second coefficient of viscosity $\secondcoefficient$, as defined in the present paper by \cref{e:secondviscosity},\cite{Rosenhead_ProceedingsRoyalSocietyLondonSeriesMathematicalPhysicalSciences_1954} has also been referred to by some authors\cite{Batchelor_1973,Sonin_2001} as the bulk viscosity; this definition results in a (misleadingly) negative value for bulk viscosity in dilute, monatomic gases.
According to the convention adopted in the present paper, bulk viscosity is $\mu_B=0$ for dilute monatomic gases, consistent with literature in acoustics and molecular kinetics.\cite{DukhinG_JChemPhys_2009}

The physical complexity of the relaxation processes modeled by bulk viscosity  contribute further to the confusion in the literature. 
Some authors\cite{Pierce_1989} attribute bulk viscosity to rotational molecular relaxation only, while others\cite{Cramer_2012} also include vibrational effects.
In the present manuscript, both the effects of vibrational and rotational molecular relaxation are incorporated within $\mu_B$ by definition. 

In acoustics, bulk viscosity effects due to rotational relaxation are of secondary importance; for air $\mubrot \approx 0.6 \mu$ for approximately all frequencies, with some dependency on temperature,\cite{Pierce_1989} as discussed later.
On the other hand, the vibrational relaxation contribution can be relatively large, e.g., $\mubvib \approx 25 \mu$ at 1000 Hz in dry air under standard atmospheric conditions. 

The combined effect of rotational and vibrational relaxation 
can be inferred from the absorption coefficient, $\alpha$---a measure of the relative wave amplitude attenuation per propagation distance---which is of particular interest in atmospheric acoustics. 
For a purely traveling tonal disturbance, spatial attenuation follows the exponential law
\begin{subequations}
	\label{e:pamp_decay}
	\begin{align}
	P_{amp} &= P_{amp,0}\,e^{\alpha \lambda \left[ -(x-x_0)/\lambda \right]}  \\ 
	&= P_{amp,0}\,e^{\alpha \lambda \left[ -(t-t_0) \omega/2\pi \right]} \, ,
	\end{align}
\end{subequations}
where $P_{amp,0}$ is the amplitude of the traveling wave at position $x=x_0$ or, equivalently, at time $t=t_0$, and $\omega$ is the angular frequency and $\lambda$ is the wavelength. 

The classical acoustic absorption coefficient, or the Stokes-Kirchoff attenuation coefficient, only accounts for thermoviscous attenuation,\cite{Temkin_1981}
\begin{align}
\label{e:classicalabsorption}
\alphacl  = \frac{\omega^2 }{2\rho_0 a_0^3} \left[ \frac{4}{3} \mu  + \frac{\left(\gamma-1\right)^2\kappa}{\gamma \Rgas  } \right] \, ,
\end{align}
where 
$\rho_0$ and $a_0$ are the base density and speed of sound, and $\kappa$ is the thermal conductivity. 
The first term on the right-hand side of \cref{e:classicalabsorption} represents the effect of viscous dissipation and the second term represents the effect of heat conduction; both processes have characteristic timescales in the gigahertz range for gases such as argon, helium, and neon.\cite{Temkin_1981}
As a result, the rate of lossy, thermoviscous momentum transfer is small, compared to momentum transfer due to ideal wave propagation.  
The classical absorption coefficient is derived from simplifying the more general form of the complex thermoviscous wave number 
of a damped monochromatic traveling wave, 
\begin{subequations}
\begin{align}
\label{e:complexwavenumber}
\alphacl & = \Imag \left[ \frac{\omega}{a_0} \sqrt{ \frac{\left(1+i\gamma\tau_{\kappa}\omega\right)\left(1+i\tau_{\nu}\omega\right)}{\left(1+i\tau_{\kappa}\omega\right)} } \right]  \\ 
& \simeq \frac{\tau_{\nu} \omega^2}{2 a_0} + \frac{\left(\gamma-1\right)\tau_{\kappa} \omega^2}{2 a_0}
\end{align}
\end{subequations}
where $\tau_{\nu} = 4 \mu/3\rho_0 a_0^2$ is the viscous relaxation time, 
$\tau_{\kappa} = \kappa / c_{p0} \rho_0 a_0^2 $ 
is the conduction relaxation time where $c_{p0}$ is the isobaric heat capacity, and the approximation is taken by neglecting 
higher-order terms in the quantities $\tau_{\nu}\omega$ and $\tau_{\kappa}\omega$.  
We note that $\tau_{\kappa}$ here should not be confused with $\tau_k$, which is the vibrational molecular relaxation time explained in \cref{sec:bv_model}.

Bulk viscosity effects are attributed to the difference between the observed attenuation rate and the rate predicted by \cref{e:classicalabsorption}. 
The total attenuation rate therefore includes contributions from rotational and vibrational molecular relaxation effects: 
\begin{align}
\label{e:absorption}
\alpha &= \alphacl + \alpha_\textrm{rot} + \sum_{k} \alpha_\textrm{vib}^{(k)}  \, ,
\end{align}
where the superscript $k$ indicates the contribution from the $k$-th species. 
For air, the species in \cref{e:absorption} are diatomic nitrogen ($k=\Ntwo$) and oxygen ($k=\Otwo$).
Other species such as water vapor do not contribute directly to \cref{e:absorption}, but instead act to adjust the vibrational relaxational frequencies. 

Combining the rotational and vibrational relaxation effects into one bulk viscosity coefficient $\mu_B$ yields the complete attenuation relation 
\begin{align}
\label{e:fullabsorption}
\alpha = \frac{\omega^2 }{2\rho_0 a_0^3} \left[ \frac{4}{3} \mu  
+  \frac{\left(\gamma-1\right)^2\kappa}{\gamma \Rgas } 
+ \mu_{B} 
\right] 
\, , 
\end{align}
which is at the core of the proposed method to estimate the value of $\mu_{B}$.

\subsection{\label{sec:intro:researchaims} Paper Outline}

The remainder of the paper is organized as follows: 
\Cref{sec:bv_model} presents the derivation of the proposed bulk viscosity model; 
\Cref{sec:computational_model} presents results from Navier--Stokes and LBM computations reproducing acoustic attenuation rates using the proposed bulk viscosity model; 
\Cref{sec:discussion} presents a discussion of the bulk viscosity coefficient as a function of pressure, temperature, frequency, and humidity and its non-dimensionalization; 
and finally, \Cref{sec:conclusion} discusses the limitations of the current approach and presents concluding thoughts.
\Cref{app:timedomain_ns} and \Cref{app:timedomain_lbm}, respectively, provide details on the Navier-Stokes and LBM implementations used.

\section{\label{sec:bv_model} Proposed Bulk Viscosity Model}

The attenuation contributions from rotational and vibrational molecular relaxation are collapsed into one bulk viscosity coefficient, such that
\begin{subequations}
\begin{align}
\label{e:mub_functional} 
\mu_B & = \mu_B(p_0,T_0,f,h_r) \\
\label{e:combinedbulkviscosity}
&=  \mubrot + \sum_k \mubvib^{(k)} \, ,
\end{align}
\end{subequations}
where $\mubrot$ is a function of only temperature $T_0$, \cite{Pierce_1989} 
and $\mubvib$ is a function of temperature, base pressure $p_0$, frequency $f$, and relative humidity $h_r$, hence yielding the general functional form of \cref{e:mub_functional}. 
This form of $\mu_B$ is sometimes referred to in the literature\cite{Emanuel_InternationalJournalEngineeringScience_1998}  as frequency-dependent bulk viscosity, as absorption of acoustic energy associated with a given frequency is the primary technique for experimentally measuring the bulk viscosity in a gas.
Such experimental measurements are in conditions of near-equilibrium, consistent with the assumptions made in the proposed model. 
In the following sections, the vibrational and rotational relaxation contributions are discussed separately.

\subsection{Vibrational bulk viscosity}

Based on the dispersion relation for plane traveling waves\cite{Pierce_1989} and statistical thermodynamics for the vibrational energy states, the attenuation contribution due to vibrational molecular relaxation for the $k$-th species is:
\begin{align}
\label{e:absorption_vibrational}
\alpha_\textrm{vib}^{(k)}  & = \frac{\left(\gamma-1\right)}{2 a_0 /\omega} \frac{c_{v,k}}{c_p} \frac{ \omega \tau_k}{1+\left(\omega \tau_k\right)^2} \\
\label{e:absorption_vibrational_cv}
c_{v,k} & = \frac{n_k}{n} \Rgas \left(\frac{\Tvibk^*}{\Tvibk} \right)^2 \exp (-\Tvibk^*/\Tvibk) \, ,
\end{align}
where 
$\tau_k$ is the vibrational molecular relaxation time, assigned according to the relaxation frequencies $f_k = \left(2\pi \tau_k\right)^{-1}$ and given for air by the semi-empirical relationships \cref{e:relaxationfrequencies:a,e:relaxationfrequencies:b},
and $c_p$ is the isobaric specific heat capacity.
Assuming quasi-equilibrium, $\Tvibk$ is set equal to the base temperature $T_0$.

\begin{figure}[tb]
	\centering
	\includegraphics[width=\onecolumnwidth]{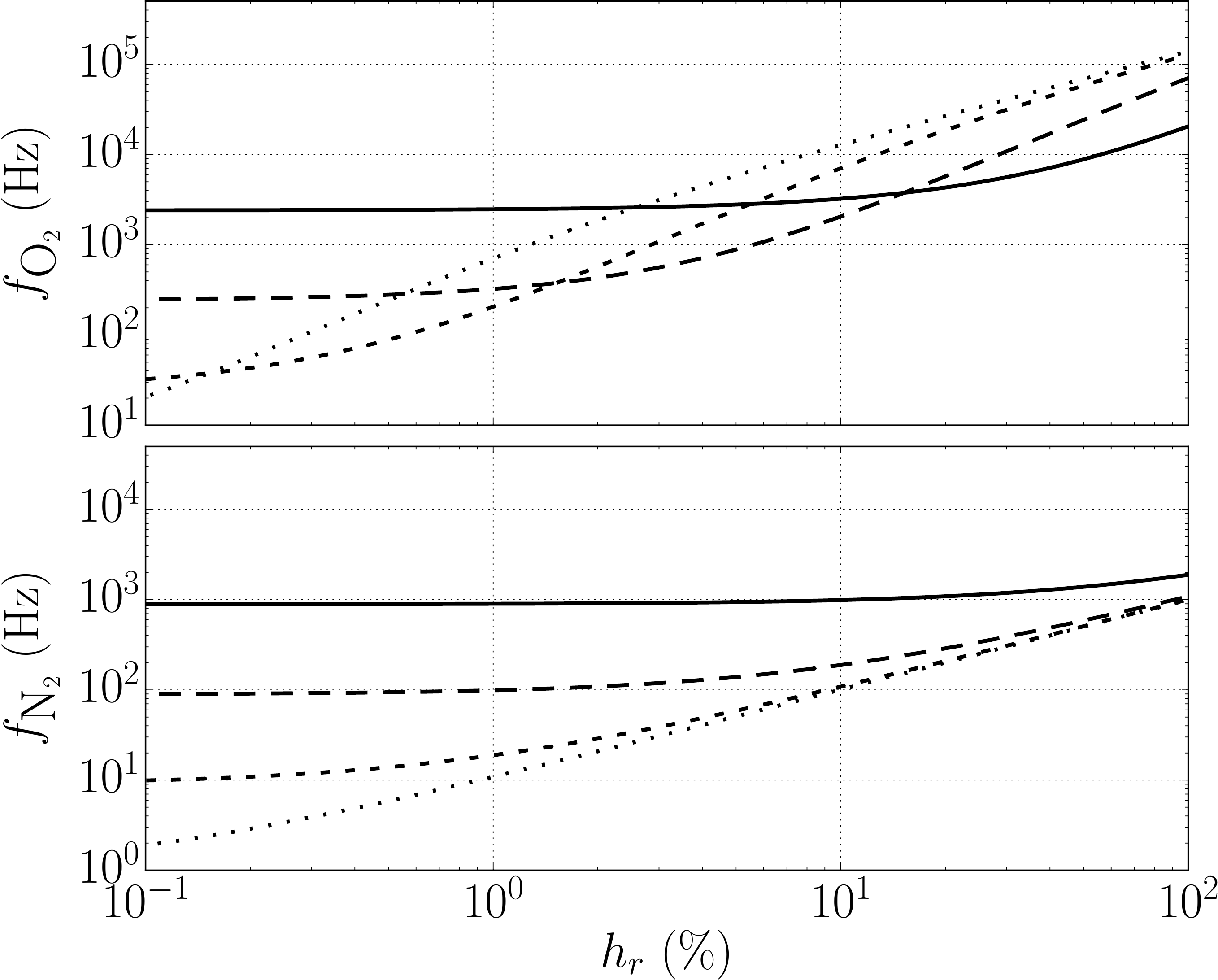}
	\caption{
	Vibrational molecular relaxation frequencies for oxygen $f_{\Otwo}=\left(2\pi \tau_{\Otwo}\right)^{-1}$ and for nitrogen $f_{\Ntwo}=\left(2\pi \tau_{\Ntwo}\right)^{-1}$ at $T_0=300$ K, plotted against relative humidity $h_r$,  for \pressurelist{}. 
	}
	\label{f:var_with_humidity}
\end{figure}

\begin{figure}[tb]
	\centering

	\includegraphics[width=\onecolumnwidth]{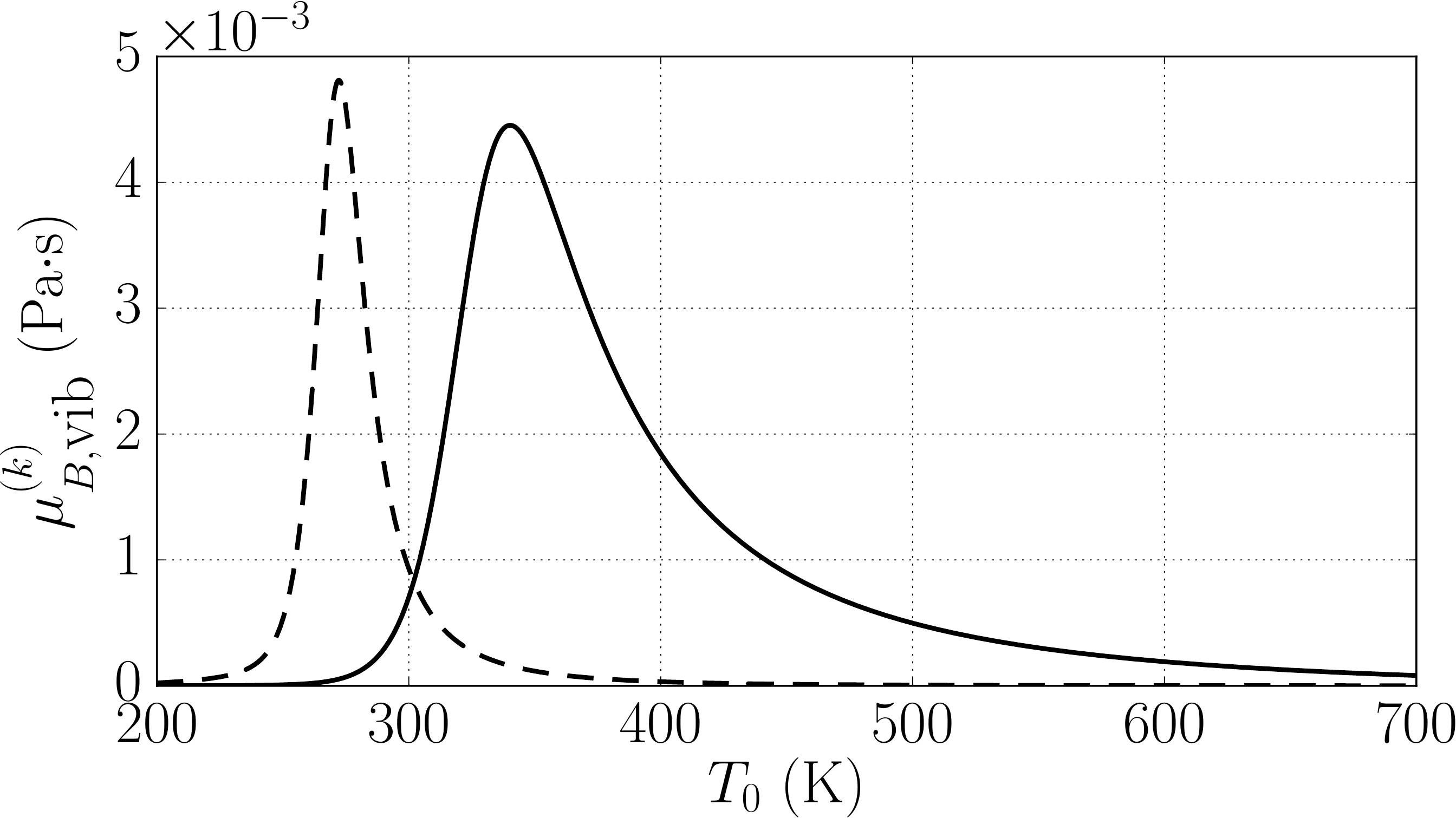}
	\caption{
		Modeled vibrational bulk viscosity $\mubvib^{(k)}$ for diatomic nitrogen $k=\Ntwo$  \legendline{}, and oxygen $k=\Otwo$ \legendlongdashed{}, plotted against temperature $T_0$ (\cref{e:vibrationrelaxation}). 
		Results presented are for $\omega/2\pi=1$ kHz in air at $h_r=20.0\%$ and $p_0=1$ atm. 
	}
	\label{f:vibbulk}
\end{figure}

\begin{table}[tb]
	\centering
\begin{tabular}{r|rr}
$k=$	& $\Otwo$ & $\Ntwo$ \\ \hline
	$n_k/n$ & \multicolumn{1}{r}{0.21} & \multicolumn{1}{r}{0.78} \\
	$\Tvibk^*$ & \multicolumn{1}{r}{2239.1 K} & \multicolumn{1}{r}{3352.0 K} \\
	\hline
	\multicolumn{1}{l|}{$\tau_k \left(h_r=0\right)$} & $6.632\cdot10^{-3}$ s & $1.789\cdot10^{-2}$ s \\
	\multicolumn{1}{l|}{	$\tau_k \left(h_r=20.0\%\right)$ } & $8.559\cdot10^{-6}$ s& $7.644\cdot10^{-4}$ s\\
\end{tabular}
	\caption{Input parameters for air and example $\tau_k$ for \cref{e:absorption_vibrational,e:absorption_vibrational_cv}. $\tau_k$ are calculated for $p_0=1\textrm{ atm}$ and $T_0=300\textrm{ K}$. 
	}
	\label{tab:air_params}
\end{table}

For air, the species involved are that of oxygen ($k=\Otwo$) and nitrogen ($k=\Ntwo$), with relevant parameters provided in \cref{tab:air_params}. 
The relaxation frequencies (in Hz) of species in air are as derived by Bass et al.\cite{BassSZBH_1995}: 
\begin{subequations} \label{e:relaxationfrequencies} \begin{align}
\label{e:relaxationfrequencies:a}
f_{\Otwo}&=\left(2\pi \tau_{\Otwo}\right)^{-1}  \nonumber \\
&= \frac{p_0}{p_{atm}} \left(24+4.04\cdot 10^4 h \frac{0.02+h}{0.391+h}\right)\\
\label{e:relaxationfrequencies:b}
f_{\Ntwo}&=\left(2\pi \tau_{\Ntwo}\right)^{-1} \nonumber \\
&= \frac{p_0}{p_{atm}} \left(\frac{T_{atm}}{T_0}\right)^{1/2} \Bigg(9+\nonumber\\&\qquad 280h\cdot \exp \left\{-4.17 \left[ \left(\frac{T_{atm}}{T_0}\right)^{1/3}-1\right]\right\}\Bigg)\\
\log_{10} & \left(p_{\textrm{sat}}/p_{atm}\right) = -6.8346 \left(T_{3p}/T_0\right)^{1.261}+4.6151 \, ,
\end{align} \end{subequations}
where 
$p_{atm} = 101325$ Pa is the reference atmospheric pressure; 
$T_{atm}=293.15$ K, the reference atmospheric temperature;
$p_{\textrm{sat}}$, the calculated saturation vapor pressure; 
and $T_{3p}=273.16$ K, the triple-point isotherm temperature.
The contribution of water vapor to the relaxation frequency of each species in air is determined via the relative humidity $h_r$, which is related to the absolute humidity $h$ as
\begin{align}
h = h_r \frac{p_\textrm{sat}}{p_0} \, .
\end{align}
Numerical constants in \cref{e:relaxationfrequencies:a,e:relaxationfrequencies:b} are dimensional. 

Sample curves for $f_{\Otwo}$ and $f_{\Ntwo}$ plotted against relative humidity are shown in \cref{f:var_with_humidity}.
While the above relations provided above are valid for air, the vibrational bulk viscosity for a different mixture can be calculated provided species-specific characteristic molecular vibration temperatures $\Tvibk^*$ and relaxation frequencies $f_k$ are known. 

Combining \cref{e:absorption,e:combinedbulkviscosity,e:absorption_vibrational,e:relaxationfrequencies}, the total vibrational bulk viscosity coefficient can be written as
\begin{align}
\label{e:vibrationrelaxation}
\mubvib &= \sum_{k}\mubvib^{(k)} \nonumber \\
\mubvib^{(k)} &=  \left(\frac{2\rho_0 a_0^3}{\omega^2}\right) \alpha_\textrm{vib}^{(k)}   \nonumber \\
&=  \bigg[\frac{p_0}{2\pi} \left(\gamma-1\right)^2 \bigg(\frac{n_k}{n}  \bigg(\frac{\Tvibk^*}{T_0} \bigg)^2 \nonumber\\
&\quad \quad \quad \quad \exp (-\Tvibk^*/T_0)\bigg) \bigg] \frac{ f_k}{f_k^2 + f^2} 
\, .
\end{align}
For dry air at standard ambient temperature and pressure at 1000 Hz,  $\mubvib \simeq 22 \mu$, 
and an example of the dependence of \cref{e:vibrationrelaxation} on temperature is shown in \cref{f:vibbulk} for air with $h_r=20$\% under standard ambient pressure.

\subsection{Rotational bulk viscosity}

The rotational molecular relaxation contribution to the overall attenuation rate, as per \cref{e:absorption,e:fullabsorption,e:combinedbulkviscosity}, is 
\begin{align}
\label{e:absorption_rotational}
\alpharot &= \frac{\omega^2 }{2\rho_0 a_0^3} \mubrot
\end{align}
where $\mubrot$ is the rotational bulk viscosity. 

\begin{figure}[tb]
	\centering
	\includegraphics[width=\onecolumnwidth]{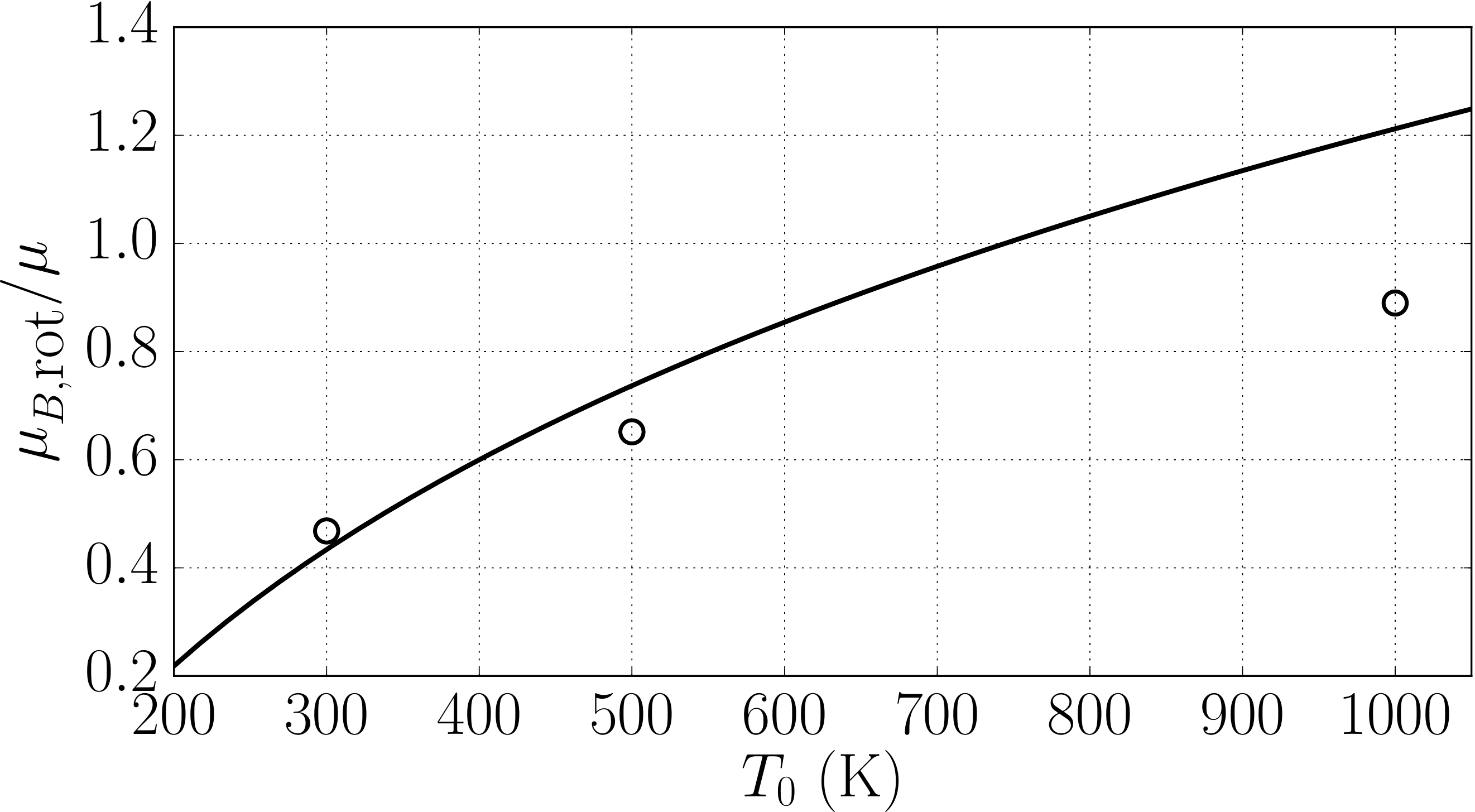}
	\caption{
		Ratio of the modeled rotational bulk viscosity and shear (or dynamic) viscosity, $\mubrot/\mu$  \legendline{}, plotted against temperature (\cref{e:rotationalviscosity,e:mu_temp_dependence}). 
		Values for the ratio $\mubrot/\mu$ by Ern and Giovangigli\cite{ErnG_EurJMechBFluids_1995} \legenddot{}. 
		Results presented are for air. 
	}
	\label{f:rotbulk}
\end{figure}

Under the ideal gas assumption, the contribution to bulk viscosity from rotational relaxation is expected to only be a function of the equilibrium temperature $T_0$. \cite{Pierce_1989,Cramer_2012}
In early literature,\cite{Pierce_1989,Thompson_1988,BassBE_1972} the rotational bulk viscosity for air is described as being between $\mubrot/\mu=0.60-0.62$, where the ratio is independent of temperature. 
However, more recent studies\cite{Emanuel_InternationalJournalEngineeringScience_1998} 
suggest that the ratio $\mubrot/\mu$ should increase with temperature.
Ern and Giovangigli\cite{ErnG_EurJMechBFluids_1995} calculated the rotational bulk viscosity via a linear systems approach by using expansion functions of the energy levels of each species and approximations for collision integrals, finding that the above ratio does in fact increase with temperature, as shown in \cref{f:rotbulk}. 

It follows from Bass et al.\cite{BassSZBH_1995,BassSZBH_1996} that the semi-empirical expression for the combined attenuation from viscous, heat conduction, and rotational bulk viscosity effects (in air) is
\begin{align}
\alphacl + \alpharot  & =  \sqrt{\frac{T_0}{T_{atm}}} \frac{p_{atm}}{p_0} c_1 f^2
\label{e:bass_combined_cl_rot}
\, ,
\end{align}
where $c_1=1.84\cdot10^{-11} \textrm{ s}^{2}\textrm{/m}$.

Combining \cref{e:bass_combined_cl_rot,e:absorption_rotational,e:absorption,e:fullabsorption} yields the final expression for rotational bulk viscosity used in this paper: 
\begin{align}
\label{e:rotationalviscosity}
\mubrot &= \frac{\gamma p_{atm}}{2\pi^2} \sqrt{\frac{\gamma \Rgas }{T_{atm}}} c_1 T_0 - \frac{4}{3} \mu - \frac{\left(\gamma-1\right)^2 \kappa }{\gamma \Rgas } \, .
\end{align}

The resulting rotational bulk viscosity 
yields $\mubrot/\mu$ ratios in good agreement with Ern and Giovangigli's estimates, similarly ranging from 0.433 to 1.21 between 300 K to 1000 K, as shown in \cref{f:rotbulk}. 
The shear viscosity $\mu$ used here is given by
\begin{align}
\label{e:mu_temp_dependence}
\mu = \mu_{\textrm{ref}}\left(T/T_\textrm{ref}\right)^{n_{\nu}}
\end{align}
where $n_{\nu} = 0.76$ is the viscosity power-law exponent and $\mu_{\textrm{ref}} =1.98\times10^{-5} \, \textrm{kg}\, \textrm{m}^{-1} \textrm{s}^{-1}$ and $T_{\textrm{ref}}=300\,\textrm{K}$ are the reference viscosity and temperature. 
Alternative formulations such as Sutherland's law\cite{Sutherland_1893} can also be used. 

The overall absorption predicted using the combined effective bulk viscosity developed in the preceding \cref{e:vibrationrelaxation,e:rotationalviscosity} is shown in \cref{f:airattenuation}; these curves accurately replicate experimental measurements of absorption in air.

\begin{figure}[tb]
	\centering
	\includegraphics[width=\onecolumnwidth]{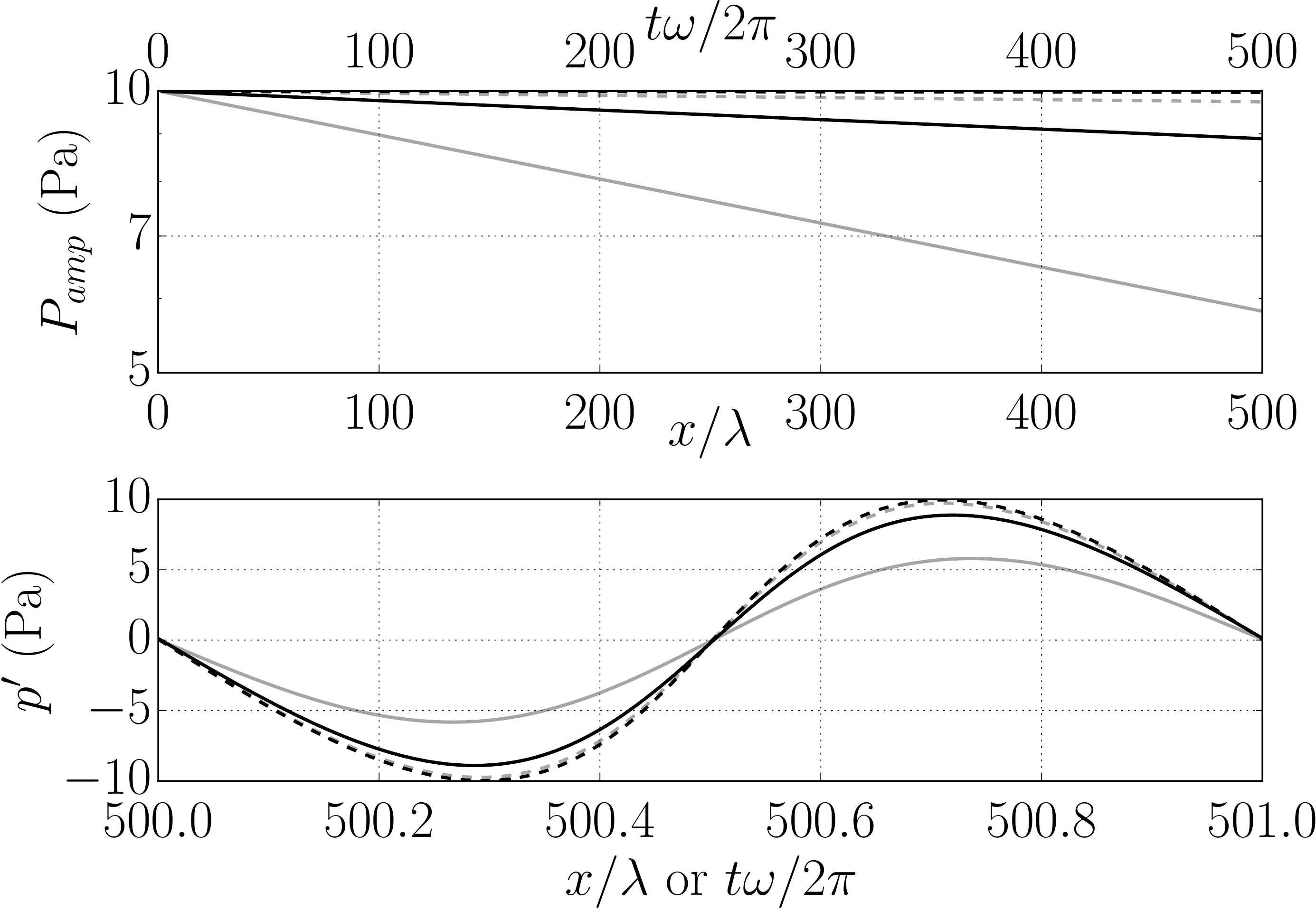}
	\caption{
		Time history of pressure amplitudes (top) of a freely-traveling wave for
		$\omega/2\pi=1$ kHz with zero bulk viscosity \legenddashed[black]{} and with effective bulk viscosity \legendline[black]{}, 
		for $\omega/2\pi=10$ kHz with zero bulk viscosity \legenddashed[gray]{} and with effective bulk viscosity \legendline[gray]{}. 
		Pressure fluctuation versus propagation distance in acoustic cycles for $x/\lambda=t\omega/2\pi>500$ (bottom). 
		Results presented are from the Navier--Stokes solver with conditions of $x_0=t_0=0$, $h_r=20\%$, $p_0=101325$ Pa, $T_0=$300 K, and $P_{amp,0}=10$ Pa.  
	}
	\label{f:attenuation_periods}
\end{figure}

\section{\label{sec:computational_model} Time-domain acoustic wave attenuation simulations}

The bulk viscosity model as developed in \cref{sec:bv_model} is coupled with both a Navier--Stokes and a lattice Boltzmann method (LBM) solver 
for verification purposes. 
Implementation details can be found in \cref{app:timedomain_ns} for the Navier--Stokes solver and in \cref{app:timedomain_lbm} for the LBM solver. 

The computational setup is identical for both solvers. 
Monochromatic planar traveling wave simulations with 4096 points per wavelength have been performed in a periodic domain. 
Frequencies in the range $f=10^1-10^5$ Hz have been tested with relative humidity levels of dry air, 20\%, and saturated air. 
Results for relative humidity level of 20\% only are shown in \cref{f:airattenuation}. 

\subsection{\label{sec:computational_model_ns}Fully compressible Navier--Stokes simulations}

In the fully compressible Navier--Stokes simulations, the governing equations for mass, momentum, and energy are solved in conservative form (see \cref{app:timedomain_ns}).

Due to the monochromatic nature of the wave propagation, a fixed value of the effective bulk viscosity has been used for any given frequency. 
Initial conditions were set as a pure isentropic traveling wave with initial pressure amplitude $P_{amp,0} = 10$ Pa. 
Numerical experiments yielded $\alpha$ as annotated in \cref{f:airattenuation}, which were extracted from time-series decaying pressure amplitudes, as shown in \cref{f:attenuation_periods}, where instantaneous pressure $p'$ and amplitude $P_{amp}$ are plotted against non-dimensional timescales as introduced in \cref{e:pamp_decay}.  
Attenuation matched to within 2\% for all tested combinations.

\subsection{\label{sec:lbm} Lattice Boltzmann simulations}

The lattice Boltzmann method is derived from Boltzmann's kinetic theory of gases and is analogous to a finite difference method for solving the Boltzmann equation. 
While often used to simulate incompressible flow problems, LBM has been successfully applied to compressible flow and acoustic problems as well, and the derivation of the compressible Navier--Stokes equations from the lattice Boltzmann equations is well established.\cite{Latt_2007}

\begin{figure*}
	\centering
	\includegraphics[width=\twocolumnwidth]{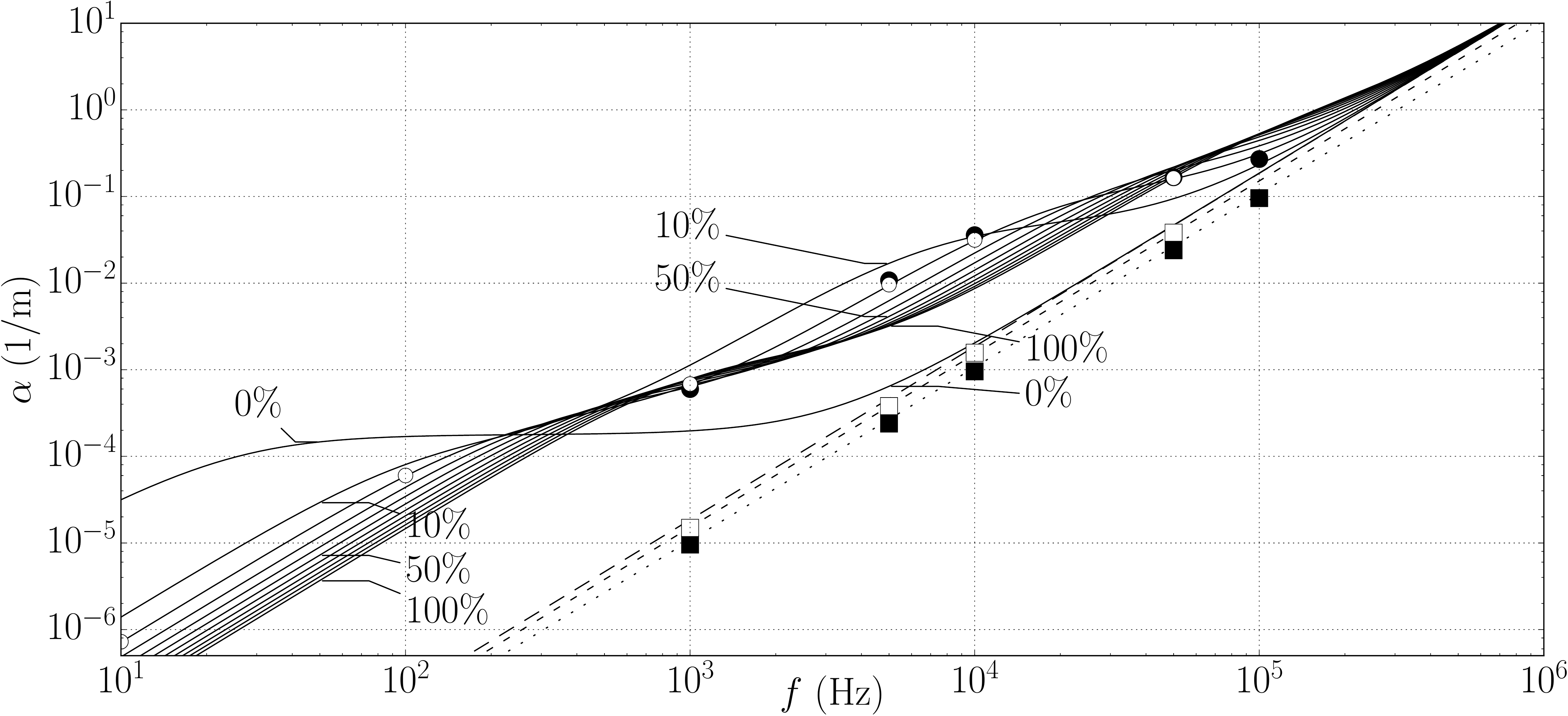}
	\caption[tikzcaption]{
		Acoustic amplitude attenuation rate (\cref{e:pamp_decay}) per unit length of propagation in air versus frequency, at $T_0=300$ K and $p_0=1$ atm, with semi-empirical expressions \legendline{} (\cref{e:absorption,e:vibrationrelaxation,e:rotationalviscosity}) and Stokes-Kirchoff attenuation expressions with rotational bulk viscosity \legendlongdashed{}, without rotational bulk viscosity \legenddashed{} , and without bulk viscosity nor conduction \legendshortdashed{} (\cref{e:classicalabsorption}); 
		relative humidity in percentage reported in figure. 
		Computationally-determined absorption via the Navier--Stokes solver for a relative humidity of $h_r=20$\% for zero bulk viscosity \legendsquare{} 
		and for calculated effective bulk viscosity \legenddot{}; 
		absorption in the lattice Boltzmann solver for a relative humidity of 20\% for zero bulk viscosity \legendsquare[black]{} 
		and for calculated effective bulk viscosity \legenddot[black]{}. 
	}
	\label{f:airattenuation}
\end{figure*}

The implementation of the proposed bulk viscosity model was carried out with a simple 2D LBM solver, using the common D2Q9 (two-dimensional, 9 velocities) scheme, with results shown in \cref{f:airattenuation}. 
The LBM solver is extended with a multiple relaxation time (MRT) model, allowing the introduction of a separate bulk viscosity parameter (see \cref{app:timedomain_lbm}). 
The MRT model was coupled with the bulk viscosity coefficient as calculated in \cref{sec:bv_model}. 

The wave attenuation rate reproduced by the LBM simulations matches the semi-empirical curves with an error of under 15\% for all tested combinations. 
Tuning of the MRT model, in particular that of the non-dimensionless viscosity and the chosen diagonalization, is specific for different fluid problems, and can reduce the observed discrepancy. 

\section{\label{sec:discussion}Discussion}

\subsection{\label{sec:bv_function} Parametric Study}

\begin{figure*}
	\centering
\includegraphics[width=\twocolumnwidth]{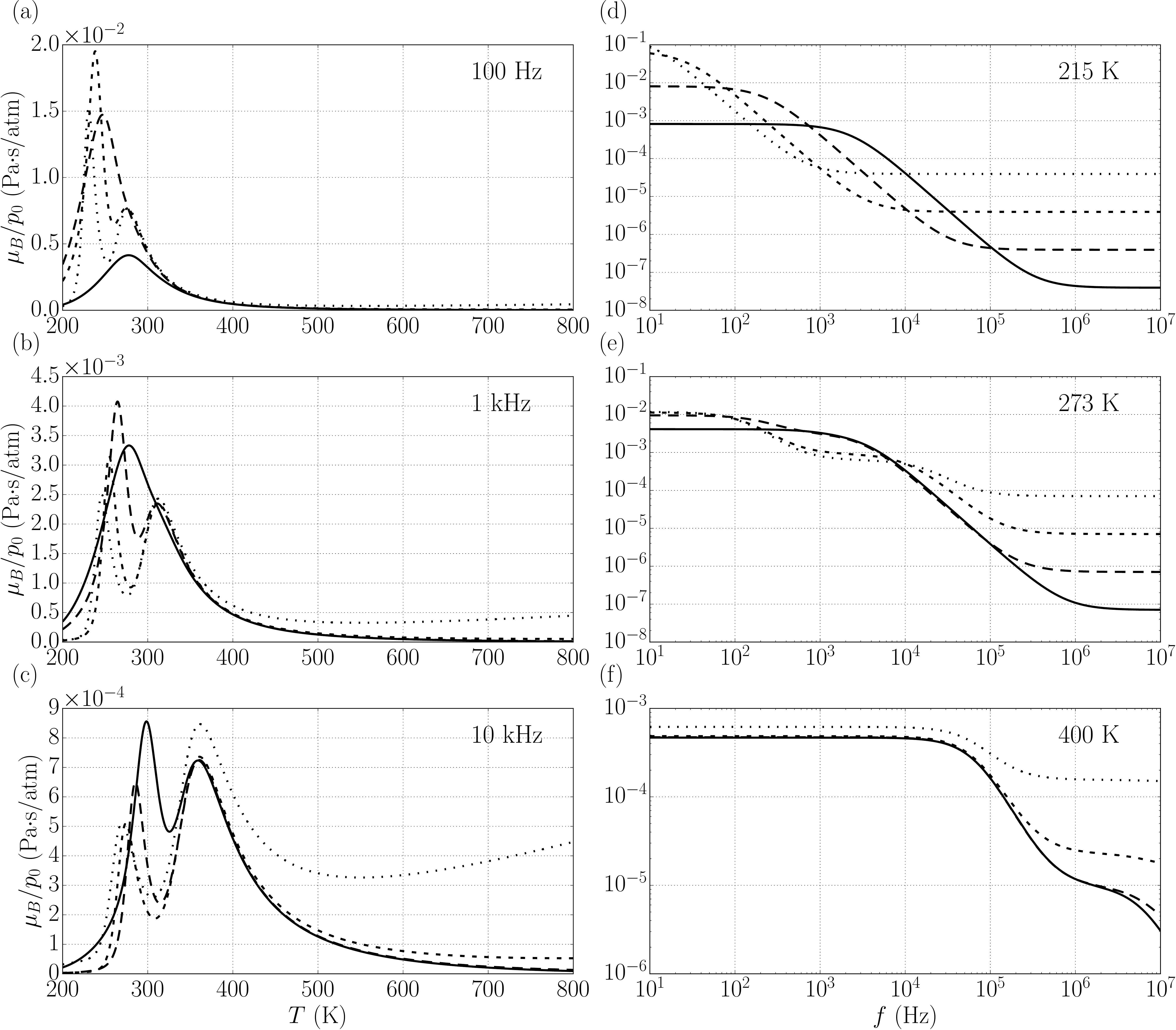}
	\caption[tikzcaption]{
		Pressure-normalized bulk viscosity $\mu_{B}/p_0$ for air at $h_r=80\%$. 
		Bulk viscosity is plotted against temperature at frequencies of 100 Hz (a), 1 kHz (b), and 10 kHz (c) with \pressurelist{}.
		Bulk viscosity is plotted against frequency at temperatures of 215.15 K (d), 273.15 K (e), and 400 K (f) with \pressurelist{}. 
	}
	\label{f:bulk_variation_notdry}
\end{figure*}

\begin{figure*}
	\centering
\includegraphics[width=\twocolumnwidth]{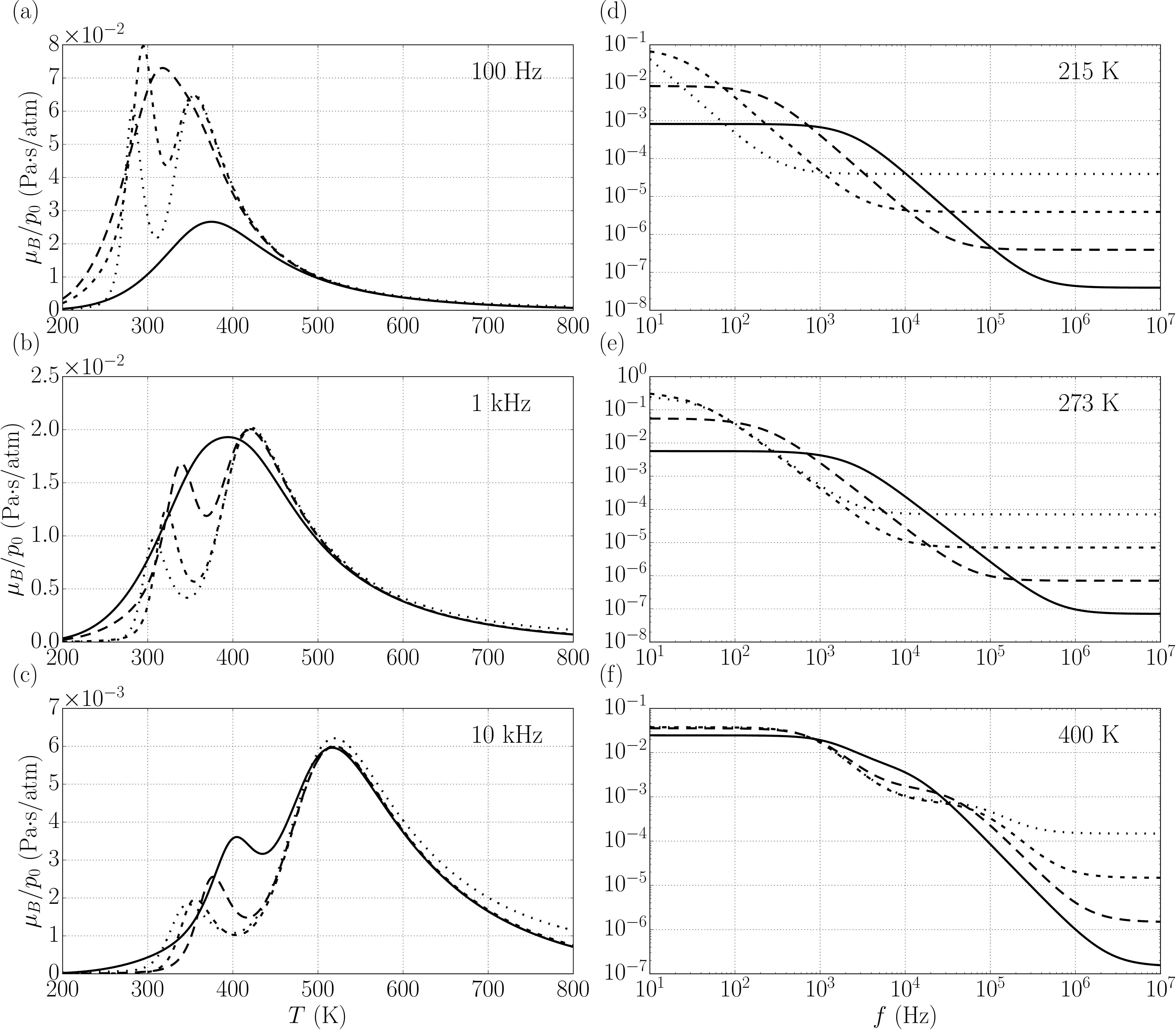}
\caption[tikzcaption]{
		Pressure-normalized bulk viscosity $\mu_{B}/p_0$ for air at $h_r=1\%$. 
	Bulk viscosity is plotted against temperature at frequencies of 100 Hz (a), 1 kHz (b), and 10 kHz (c) with \pressurelist{}.
	Bulk viscosity is plotted against frequency at temperatures of 215.15 K (d), 273.15 K (e), and 400 K (f) with \pressurelist{}.
}
	\label{f:bulk_variation_hr_low}
\end{figure*}

\begin{figure*}
	\centering
\includegraphics[width=\twocolumnwidth]{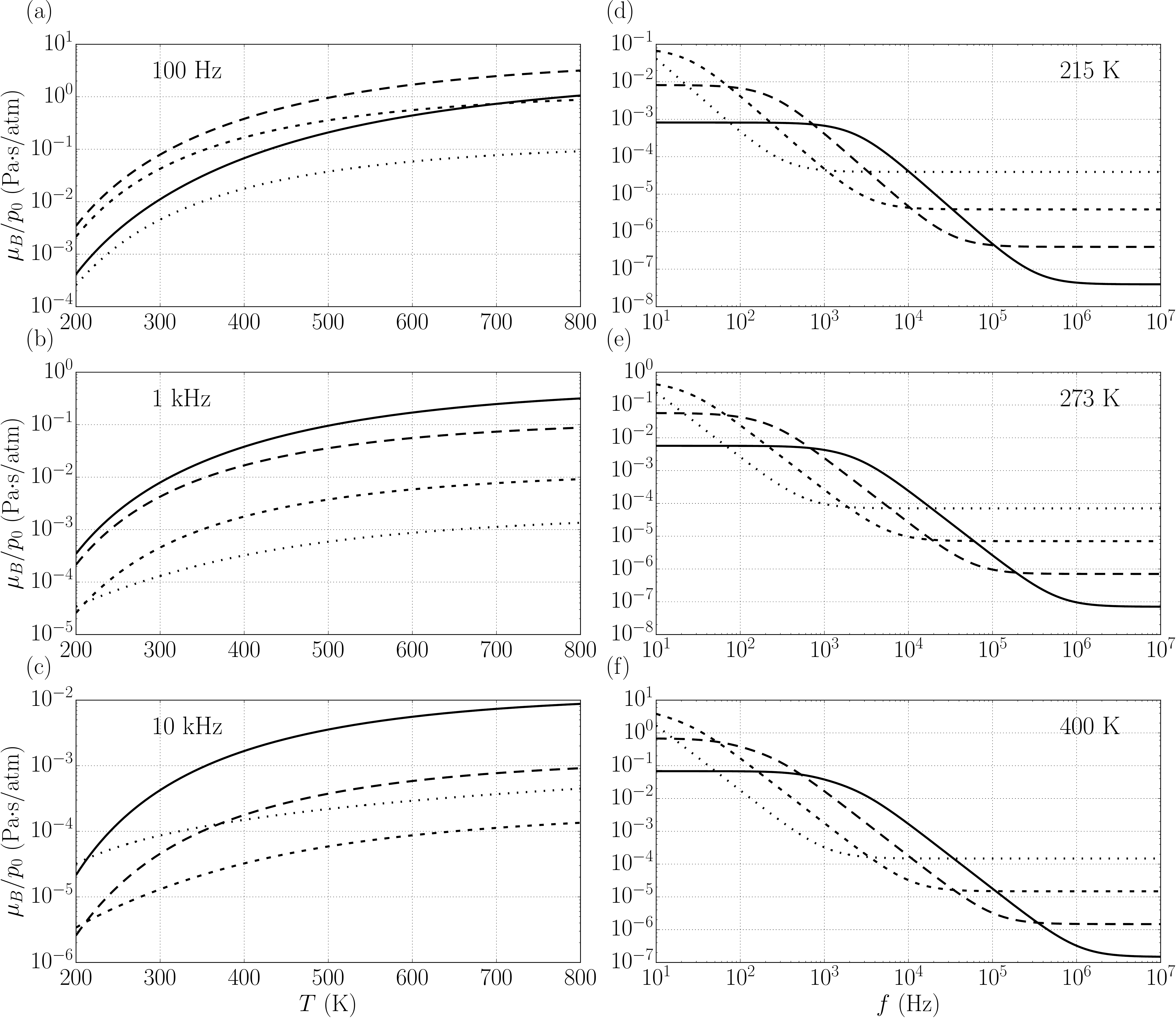}
	\caption[tikzcaption]{
		Pressure-normalized bulk viscosity $\mu_{B}/p_0$  for dry air. 
		Bulk viscosity is plotted against temperature at frequencies of 100 Hz (a), 1 kHz (b), and 10 kHz (c) with \pressurelist{}.
		Bulk viscosity is plotted against frequency at temperatures of 215.15 K (d), 273.15 K (e), and 400 K (f) with \pressurelist{}.
	}
	\label{f:bulk_variation_dry}
\end{figure*}

In this section, we explore the functional dependency of \cref{e:mub_functional} for air only.  
The dependence of $\mubvib$ and $\mubrot$ on temperature, pressure, and frequency is shown in \cref{f:bulk_variation_notdry} for air with a relative humidity of $h_r=80\%$, \cref{f:bulk_variation_hr_low} for air with a relative humidity of $h_r=1\%$, and \cref{f:bulk_variation_dry} for dry air. 
The bulk viscosity is plotted as a sum of both nitrogen and oxygen species contributions. 

Several trends are evident. 
First, the value of bulk viscosity is larger at low frequencies; while this does not necessarily mean that overall attenuation rate is larger at low frequencies, it does suggest that bulk viscosity effects cannot be completely neglected at low frequencies (e.g., infrasonic atmospheric wave propagation). 
Second, the bulk viscosity value increases with pressure at lower frequencies, noting the plot normalization with respect to the base pressure. 
Third, peaks, which at low pressures can be identified as being due to either nitrogen or oxygen vibrational frequencies, tend to merge  at higher pressures, as illustrated by figures \ref{f:bulk_variation_notdry}a and \ref{f:bulk_variation_notdry}b. 

When plotted against frequency, the vibrational bulk viscosity decreases exponentially beyond the vibrational relaxation frequency of each species, as seen in subplots (d-f) of  \cref{f:bulk_variation_notdry,f:bulk_variation_hr_low,f:bulk_variation_dry}.
At higher frequencies, the bulk viscosity contribution from rotational relaxation defines the minimum value of $\mu_B$. 

\begin{figure}[tb]
	\centering
	\includegraphics[width=\onecolumnwidth]{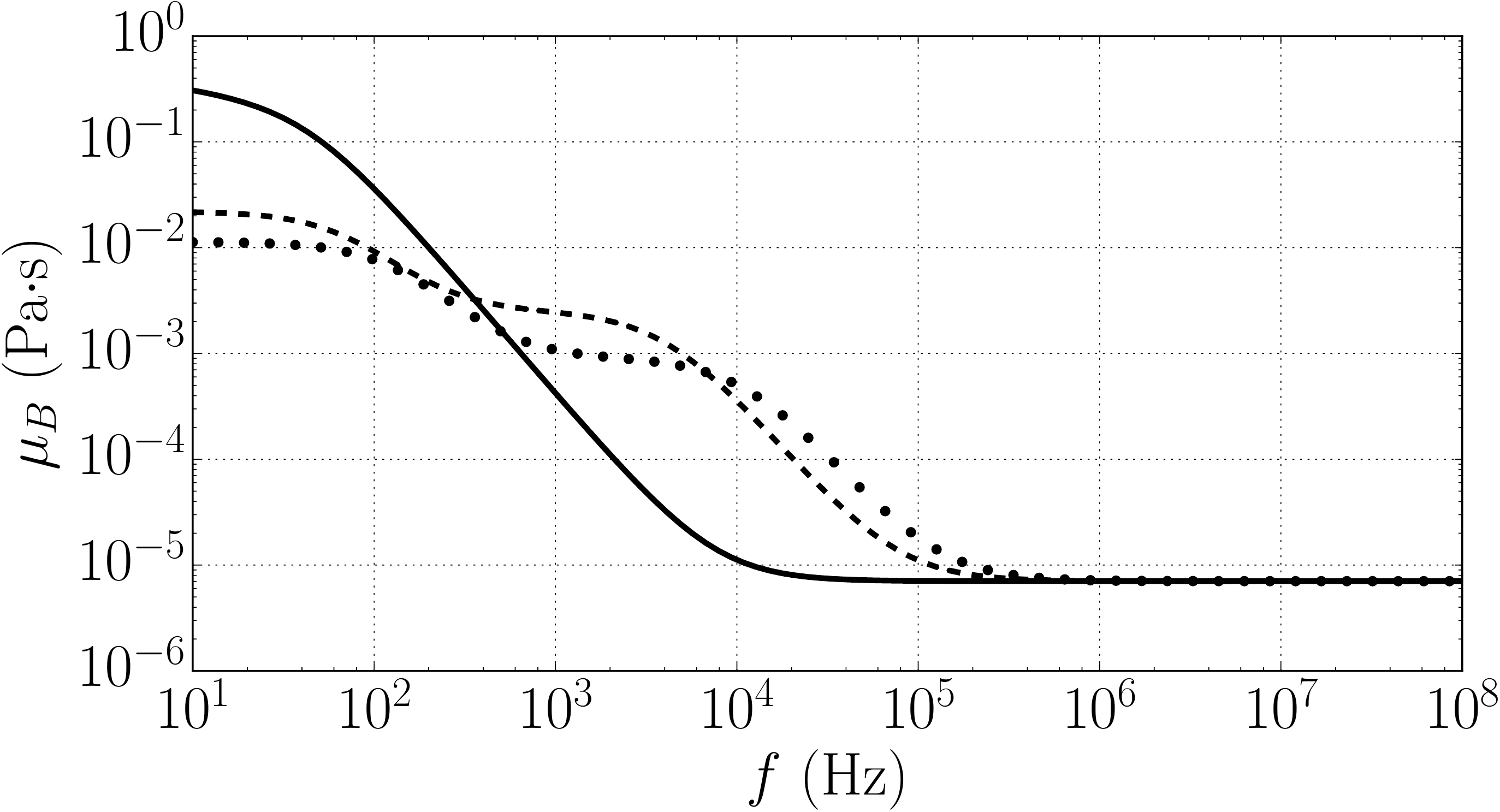}
	\caption{
		Bulk viscosity relative to frequency as calculated for air at $T_0=273.15$ K and $p_0=1.0$ atm for relative humidity levels of $h_r=1$\% \legendline{}, 40\% \legenddashed{}, and 80\% \legenddotSmultiple[black]{}. 
	}
	\label{f:mu_vary_humidity}
\end{figure}

The presence of water vapor drastically affects the magnitude of the vibrational bulk viscosity, as seen in \cref{f:bulk_variation_notdry,f:bulk_variation_hr_low,f:bulk_variation_dry}. 
Reducing humidity from $h_r=80\%$ to $1\%$ increases the maximum bulk viscosity value across the considered temperature and pressure combinations, and also shifts the peak towards higher temperatures. 
This shift demonstrates the need to evaluate bulk viscosity as a function of humidity, temperature, as well as pressure. 
As a numerical example, at $T_0=273.15$ K, reducing humidity from 80\% to 1\% increases bulk viscosity for frequencies approximately below 500 Hz, but decreases bulk viscosity at mid frequency ranges (up through 20--30 kHz) and at higher frequency ranges (above 30 kHz).
Increasing humidity increases the relaxation frequency of both nitrogen and oxygen species, and as seen in \cref{e:vibrationrelaxation}, bulk viscosity will, for each species, reach a maximum at a different combination of pressure and temperature.
Because oxygen has a higher relaxation frequency than that of nitrogen, there is a region between the two frequencies for which $\mu_B$ achieves a maximum relative to relative humidity; 
at lower frequencies, bulk viscosity decreases as humidity increases, and at higher frequencies, bulk viscosity increases as humidity increases, as seen in \cref{f:mu_vary_humidity}. 

Bulk viscosity in air with zero relative humidity (\cref{f:bulk_variation_dry}) demonstrates no peaks with respect to temperature, as would be expected from \cref{e:relaxationfrequencies,e:vibrationrelaxation}. 
An increase in base temperature reduces vibrational relaxation frequencies for the nitrogen species, but has no effect on the oxygen species.

\subsection{Dimensionless scaling of bulk viscosity}

\subsubsection{Effective acoustic pressure}

\begin{figure}[tb]
		\begin{center}
			$(a)_{\includegraphics[width=\onecolumnwidth]{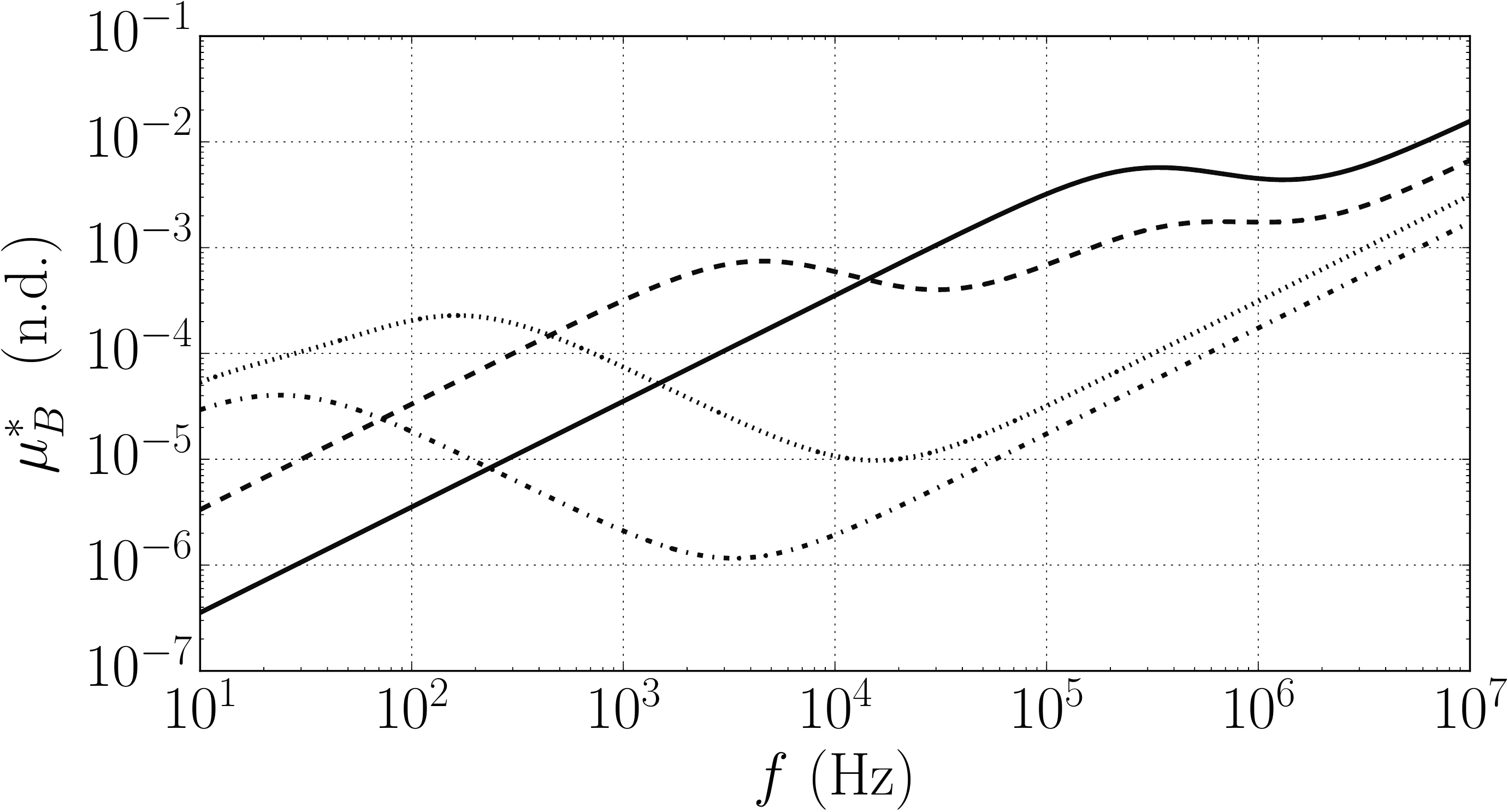}}$ 
			$(b)_{\includegraphics[width=\onecolumnwidth]{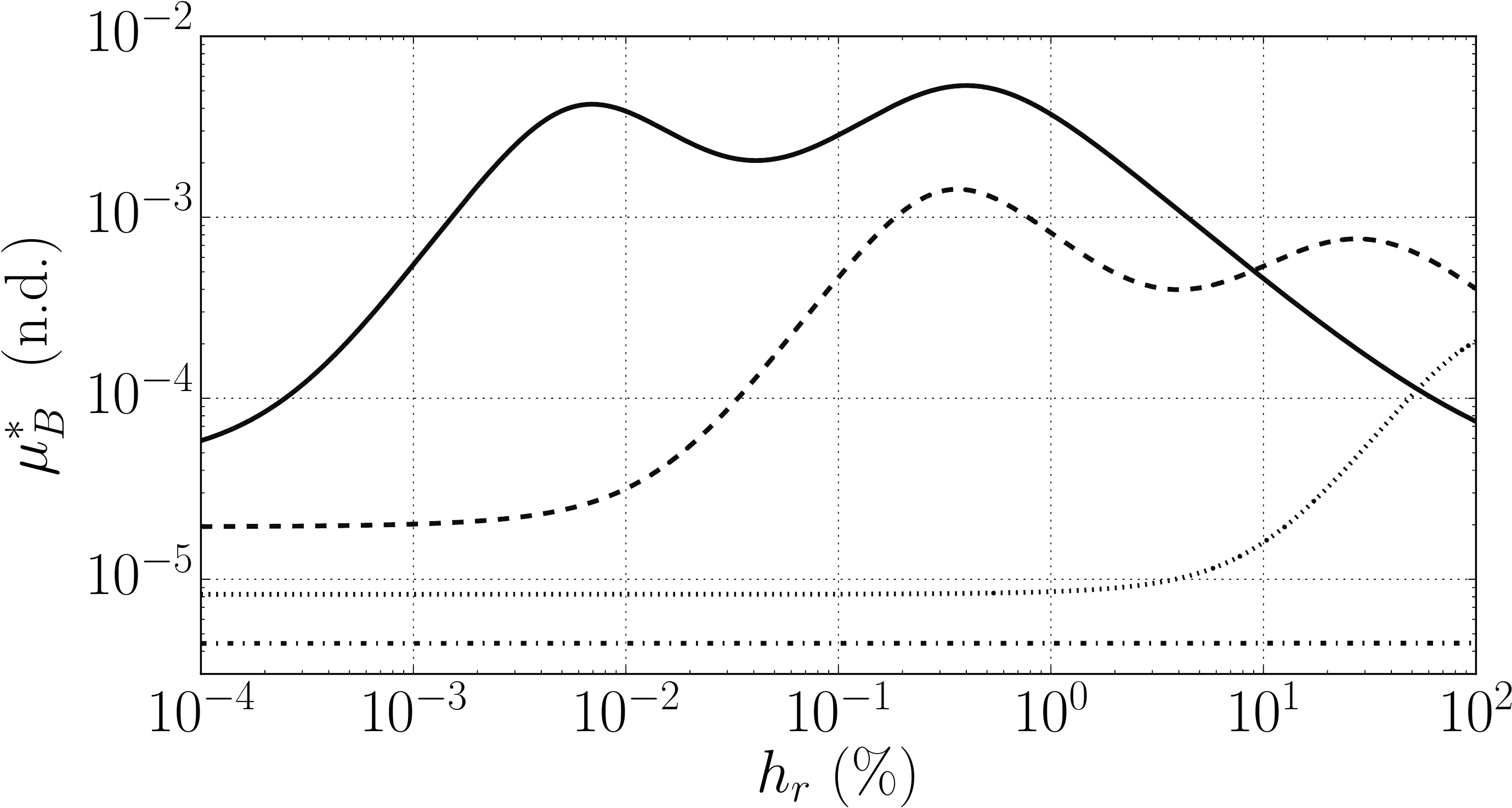}}$ 
		\end{center}

	\caption[tikzcaption]{
		Dimensionless effective bulk viscosity $\mu_B^*$ plotted against frequency at $T_0=$ 215 K \legenddasheddotted{}, 273.15 K \legendverts{}, 400 K \legenddashed{}, and 600 K \legendline{}. 
		Results shown for $h_r=5\%$ and atmospheric pressure (a). 
		Dimensionless effective bulk viscosity $\mu_B^*$ plotted against relative humidity, $h_r$, at $T_0=$ 215 K \legenddasheddotted{}, 273.15 K \legendverts{}, 400 K \legenddashed{}, and 600 K \legendline{}. 
		Results shown for $\omega/2\pi=25$ kHz and atmospheric pressure (b). 
	}
	\label{f:bulk_nd_with_temp}
\end{figure}

Effects related to bulk viscosity are largest near vibrational energy ``peaks,'' where the acoustic frequency approaches the natural frequency of a vibrational mode. 
In order to evaluate the effect of bulk viscosity under different conditions, we consider its effect on the acoustic effective (or mechanical) pressure.
Bulk viscosity effects within the momentum and energy equations (\cref{eq:navierstokes}) result in an adjustment to the thermodynamic pressure, 
\begin{align}
\label{e:momentumbulkviscosity}
p_{\textrm{eff}} &= p - \mu_{B} \nabla \cdot \vec{u}' \, ,
\end{align}
an expression which absorbs $\mu_B$ into the effective pressure gradient term. 
Per linear acoustics and assuming idealized oscillations, the continuity equation is
\begin{align}
\frac{1}{\rho_0 a_0^2} \frac{\partial p'}{\partial t} + \nabla \cdot \vec{u'} &= 0 \, . 
\end{align}
Adopting the harmonic convention $p'\sim\hat{p} e^{i\omega t}$ and assuming either standing-wave or traveling-wave phasing (both corresponding to conditions where pressure oscillations lead the velocity divergence term $\nabla \cdot \vec{u}'$ by $90^{\circ}$), the above equation simplifies to 
\begin{align}
\label{e:continuityfourier}
\frac{\omega}{\rho_0 a_0^2} \left| \hat{p} \right| - \left|\widehat{ \nabla \cdot \vec{u'}  }\right| &= 0  \, . 
\end{align}
Finally, combining \cref{e:momentumbulkviscosity,e:continuityfourier} results in 
\begin{align}
\frac{\mu_{B} \omega}{\gamma p_0} \left| \hat{p} \right| = \mu_B \left|\widehat{ \nabla \cdot \vec{u'}  }\right| = \left| \widehat{p_{\textrm{eff}}-p}\right|  \, ,
\end{align}
thus suggesting the dimensionless group 
\begin{align}
\mu_B^* = \frac{\mu_{B} \omega}{\gamma p_0}= \frac{\mu_{B} \left|\widehat{ \nabla \cdot \vec{u'}  }\right|}{\left| \hat{p} \right|} = \frac{\left| \widehat{p_{\textrm{eff}}-p}\right|}{\left| \hat{p} \right|}
 \, , 
\label{e:nd_mub_star}
\end{align}
where $\mu_B^*$ 
is a measure of the relative impact of bulk viscosity on pressure fluctuations. 

Traditional attenuation curves, such as \cref{f:airattenuation}, are measured relative to attenuation per meter and tend to belie the effect of bulk viscosity at high frequencies. 
When measured relative to acoustic wavelength, the bulk viscosity contribution to attenuation can be as large as 1\% of pressure amplitude and has a magnitude peak varying with gas temperature, pressure, and humidity, as shown in \cref{f:bulk_nd_with_temp}. 

\subsubsection{Dimensionless groups}
\label{sss:dimensionlessgroups}

For many fluid problems, dimensionless groups can reduce the number of variables involved, helping to understand the underlying phenomenon and simplifying experiments. 
However, the proposed bulk viscosity model does not offer a straightforward non-dimensionalization, as the 
direct application of the Buckingham $\pi$ theorem to \cref{e:vibrationrelaxation,e:rotationalviscosity,e:combinedbulkviscosity} is limited by constants in the relaxation frequencies (\cref{e:relaxationfrequencies}) and by the saturation vapor pressure. 

As a result, for the general air case, neither frequency $f$ nor relative humidity $h_r$ can be part of a useful dimensionless group. 
Furthermore, temperature $T_0$ is constrained by the triple-point isotherm temperature $T_{3p}$ and pressure $p_0$ is similarly constrained due to the saturation pressure, $p_{sat}$, which depends on $T_0$.

\begin{figure}[tb]
	\begin{center}
		\includegraphics[width=\onecolumnwidth]{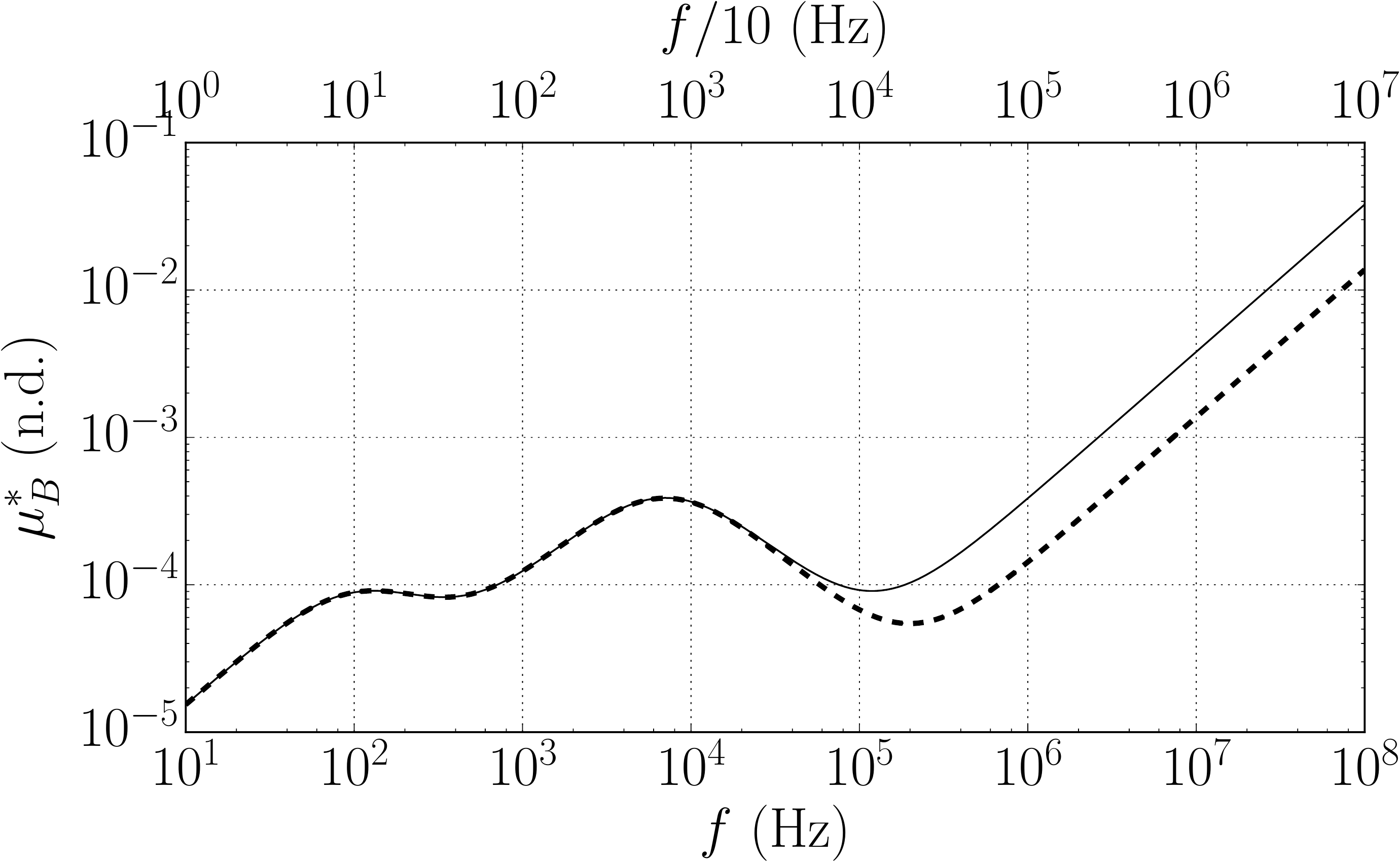}
	\end{center}
	\caption[tikzcaption]{
		Dimensionless bulk viscosity $\mu_B^*$ for air versus frequency at $T_0=$ 300 K for conditions of unscaled $f$, $p_0=1$ atm, $h_r=10$\%, and $\gamma=1.4$ \legendline{} and of scaled dimensionless bulk viscosity (see \cref{e:vib_nd_scaling}) for conditions of scaled frequency $f/10$ kHz, $p_0=0.1$ atm, $h_r=1$\%, and $\gamma=1.66$ \legenddashed{}. 
		This corresponds to scaling parameters of $b_1=0.1$ and $b_2=1.19$, such that $b_2 \gamma=1.66$. 
		Under conditions of $b_2=1$, the plots would be identical. 
	}
	\label{f:bulk_nd_scaling}
\end{figure}

To find useful scaling parameters, we consider two special cases. 
In the first case, relative humidity $h_r$ is allowed to vary without bounds. 
Combining \cref{e:vibrationrelaxation,e:rotationalviscosity} suggests that the dimensionless group $\mu_B^*$ remains constant so long as scaling is applied simultaneously to frequency, pressure, and relative humidity, 
\begin{align}
\label{e:scalinglaw1}
\mu_{B}^* = \frac{\mu_{B} \omega}{\gamma p_0} & = \textrm{fn} \left(f_0, p_0, h_r, T_0 \right) \nonumber  \\
& = \textrm{fn} \left(b_1 f_0, b_1 p_0, b_1 h_r, T_0 \right)  \, ,
\end{align}
for positive scaling parameter $b_1$. 

The second case considers splitting the dimensionless bulk viscosity into rotational and vibrational relaxational components. 
The rotational bulk viscosity, as noted before, solely depends on temperature, and the effect associated with rotational bulk viscosity may be neglected at lower frequencies. 
Considering instead only the vibrational bulk viscosity, a dimensionless vibrational bulk viscosity 
\begin{align}
\mubvib^* = \frac{\sum_k \mubvib^{(k)} \omega}{\gamma p_0}
\end{align}
not only follows the scaling law \cref{e:scalinglaw1} but also 
remains constant if the heat capacity ratio $\gamma$ is scaled as
\begin{align}
\mubvib^* & = \textrm{fn} \left(f_0, p_0, h_r, T_0, \gamma \right) \nonumber \\
& = \left(b_2  \frac{\left(\gamma-1\right)^2}{\left(b_2 \gamma-1\right)^2}\right) \textrm{fn} \left(b_1 f_0, b_1 p_0, b_1 h_r, T_0, b_2  \gamma \right)  \, , 
\label{e:vib_nd_scaling}
\end{align}
where both $b_1$ and $b_2$ are arbitrary scaling parameters. 
For \cref{e:vib_nd_scaling}, $\mubvib^*\approx\mu_{B}^*$ for small $f$, as shown in \cref{f:bulk_nd_scaling}. 
Because $\gamma$ is related to the degrees of freedom of the fluid in question, it is unlikely that scaling with $b_2$ can be used without disrupting the general model. 

An example of both forms of scaling is shown in \cref{f:bulk_nd_scaling}. 
Because the vibrational bulk viscosity has a dominant effect in the low-frequency regime, the scaling of $\mu_B^*$ according to \cref{e:vib_nd_scaling} differs by less than 1\% for frequencies under 10 kHz. 

\subsubsection{Self-similarity of wave attenuation}

\begin{table}[tb]
	\centering
	    \begin{tabular}{r|rrr}
		& \multicolumn{1}{c}{A} & \multicolumn{1}{c}{B} & \multicolumn{1}{c}{C} \\ \hline
$f$ (kHz) & 0.5 & 1 & 2 \\
$p_0$ (atm) & 0.5 & 1 & 2 \\
$h_r$ (ref.) & 10\% & 10\% & 10\% \\
		& \tikztriangle[white]{}   & \tikzsquare[white]{}  & \tikzpentagon[white]{} \\
$h_r$ (similitude) & 5\% & 10\% & 20\% \\
		& \tikztriangle{}   & \tikzsquare{}  & \tikzpentagon{} \\
		\hline
	\end{tabular}
	\caption{Input parameters for self-similar simulations, as used in \cref{f:selfsimilarity}. 
	}
	\label{tab:selfsimilarity}
\end{table}

\begin{figure}[tb]
	\centering
	\includegraphics[width=\onecolumnwidth]{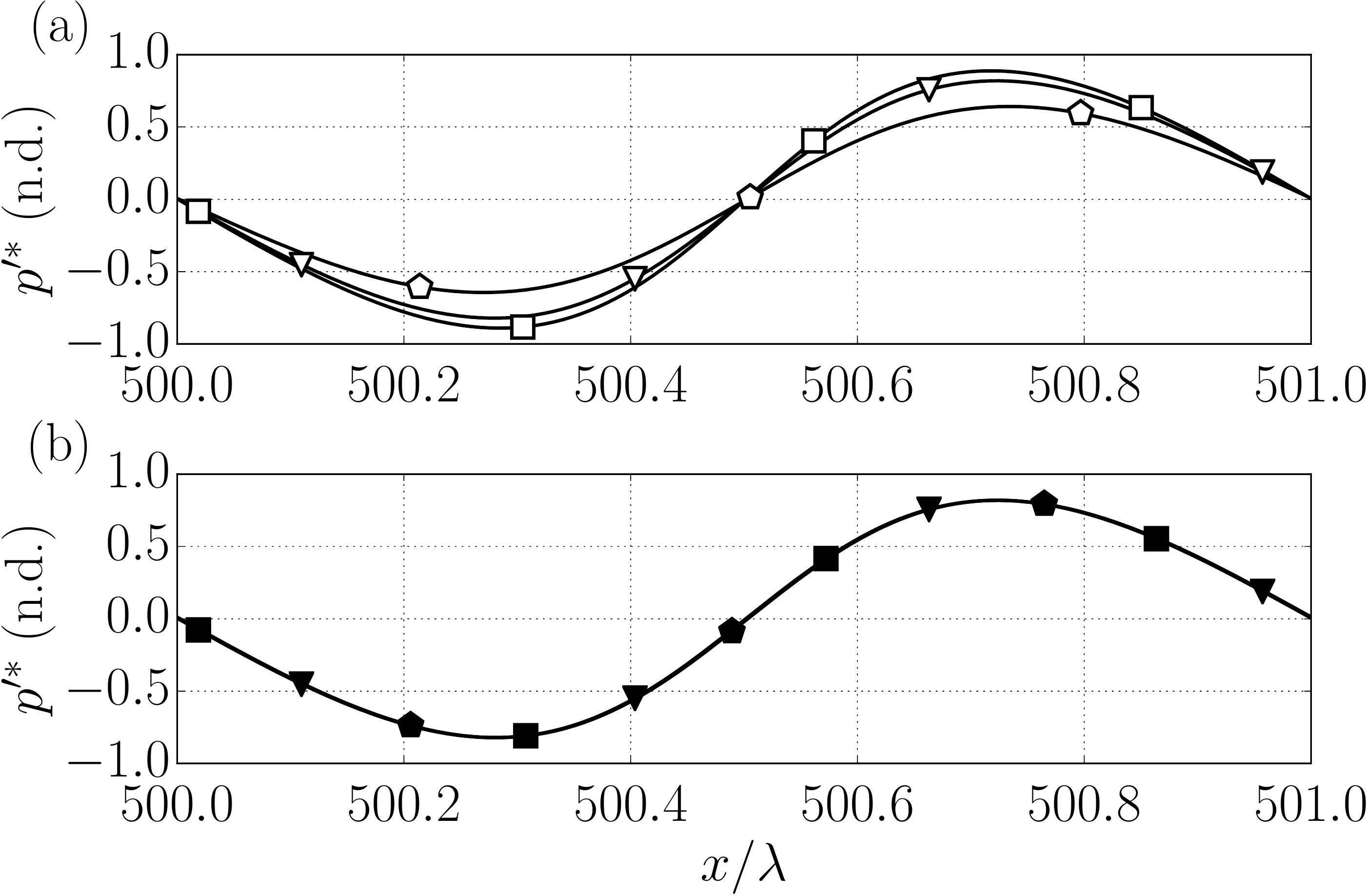}
	\caption{
		Scaled dimensionless pressure fluctuations $p'^* = p' / P_{amp,0}$ from Navier--Stokes simulations of a freely-traveling wave for acoustic cycle $x/\lambda>500$ (see \cref{e:dimensionlesscollapse,e:dimensionlessinputs})
		for cases in \cref{tab:selfsimilarity}: pressure-frequency similitude scaling (a) and pressure-frequency-humidity scaling (b). 
	}
	\label{f:selfsimilarity}
\end{figure}

A dimensionless collapse of the spatial attenuation relationships \cref{e:pamp_decay,e:fullabsorption} gives 
\begin{align}
\label{e:dimensionlesscollapse}
\log P_{amp}^* &= \pi \left[\mu^* +\kappa^* + \mu_B^*\right] x^*
\end{align}
where 
\begin{align}
P_{amp}^* &= P_{amp} /  P_{amp,0} &
x^* &= \left(x_0-x\right)/\lambda \nonumber \\ 
\mu^* &= \frac{\omega}{\gamma p_0} \frac{4}{3} \mu &
\kappa^* &= \frac{\omega}{\gamma p_0} \frac{\left(\gamma-1\right)^2 \kappa }{\gamma \Rgas} 
\label{e:dimensionlessinputs}
\end{align}
are the chosen normalizations and $\mu_B^*$ is as given in \cref{e:nd_mub_star}. 

The similitude scaling in \cref{e:scalinglaw1} is tested using the Navier--Stokes simulations, with both imperfect pressure-frequency scaling and pressure-frequency-humidity scaling, as reported in \cref{tab:selfsimilarity}. 
The simulation results are shown in \cref{f:selfsimilarity}; as expected, the scaling form of \cref{e:scalinglaw1} produces self-similar attenuation rates if the simulations account for the effect of humidity in air.

\section{\label{sec:conclusion} Conclusion}

Bulk viscosity is often neglected in fluid problems, for lack of an established model. 
We have presented a bulk viscosity model valid for tonal acoustic wave propagation in air, with verification via companion compressible Navier--Stokes simulations and lattice Boltzmann method simulations. 
The bulk viscosity model captures acoustic attenuation which otherwise is severely underestimated if Stokes' hypothesis is assumed, and can be a simple addition to both time-domain and frequency-domain solvers. 
The importance of incorporating bulk viscosity depends on the fluid problem: 
attenuation related to bulk viscosity is significant at high frequencies when normalized to absolute length scales; at low frequencies, attenuation is significant when normalized by wavelength. 

The present model has several limitations: 
(1) the frequency-dependency of $\mu_B$ makes the model unsuitable for broadband time-domain simulations; 
(2) it cannot capture broadband wave dispersion attributed to  $\mu_B$; and 
(3) the current modeling framework and the ideal gas assumption can be invalid for dense gases---in such conditions, even monatomic gases can exhibit bulk viscosity effects.\cite{Bhatia_1985}

However, the present model provides several benefits, as it is: 
(1) algorithmically simple; 
(2) extensible to any gaseous mixture provided attenuation measurements and species-specific characteristic molecular vibration temperatures $\Tvibk^*$ and relaxation frequencies $f_k$ are known; and 
(3) applicable to frequency-domain analysis, commonly used in acoustics, and to near-monochromatic time-domain problems. 

The relationship between bulk viscosity, pressure, and dilatation is used to construct a bulk viscosity dimensionless group $\mu_B^*$, applicable for both standing- and traveling-wave acoustics problems. 
Non-dimensional analysis of bulk viscosity suggests that it is a nontrivial function of fluid properties, with limited opportunities to apply similitude. 
Nevertheless, the bulk viscosity dimensionless group does follow pressure-frequency similitude, with the caveat that relative humidity must be simultaneously varied with pressure and frequency. 
Because of saturation humidity, such scaling cannot always be possible. 

The difference between effective and thermodynamic pressure suggests that the effects of bulk viscosity can be manipulated in situations for which the phasing of pressure and dilatation can be decoupled. 
Isolating the effects of attenuation due to bulk viscosity from that of shear viscosity can allow for the optimization of geometries where loss is important.

\begin{acknowledgments}
Jeffrey Lin and Carlo Scalo acknowledge the support of the Inventec Stanford Graduate Fellowship and the Precourt Energy Efficiency Center Seed Grant at Stanford University. The authors also acknowledge the generous computational allocation provided to Dr. Scalo on Purdue's latest supercomputing architecture, Rice, and the technical support of Purdue's Rosen Center for Advanced Computing (RCAC). Dr. Scalo acknowledges the support of the Air Force Office of Scientific Research (AFOSR) grant FA9550-16-1-0209 and the very fruitful discussions with Dr.~Ivett Leyva (AFOSR) on ultrasonic wave attenuation in hypersonic boundary layers.
\end{acknowledgments}

\appendix

\section{\label{app:timedomain} Bulk Viscosity Implementation in Time-Domain Simulations}

\subsection{\label{app:timedomain_ns} Fully compressible Navier--Stokes}

The conservation equations for mass, momentum, and energy solved in the fully compressible Navier--Stokes simulations are, respectively, 
\begin{subequations}
	\label{eq:navierstokes}
	\begin{align}
	\frac{\partial}{\partial t} \left(\rho\right) &+ \frac{\partial}{\partial x_j} \left(\rho u_j \right)  = 0
	\label{subeq:ns1}
	\\
	\frac{\partial}{\partial t} \left(\rho u_i\right) &+ \frac{\partial}{\partial x_j} \left(\rho u_i u_j\right)  =  -\frac{\partial}{\partial x_i} p  +
	\frac{\partial}{\partial x_j} \tau_{ij}
	\label{subeq:ns2}
	\\
	\frac{\partial}{\partial t} \left(\rho \, E\right) &+ \frac{\partial}{\partial x_j} \left[ u_j \left(\rho \, E + p \right) \right] =
	\frac{\partial}{\partial x_j } \left(u_i \tau_{ij} - q_j\right)
	\label{subeq:ns3}
	\end{align}
\end{subequations}
where $x_1$, $x_2$, and $x_3$ (equivalently, $x$, $y$, and $z$) are axial and cross-sectional coordinates, $u_i$ are the velocity components in each of those directions, and $p$, $\rho$, and $E$ are respectively pressure, density, and total energy per unit mass. 
The gas is assumed to be ideal, with equation of state $p= \rho \,\Rgas\, T$ and a constant ratio of specific heats, $\gamma$. 
The gas constant is fixed and calculated as $\Rgas = p_{\textrm{ref}} \left(T_{\textrm{ref}}\,\rho_{\textrm{ref}}\right)^{-1}$, 
based on the reference thermodynamic density $\rho_{\textrm{ref}}$, 
pressure $p_{\textrm{ref}}$, 
and temperature $T_{\textrm{ref}}$.
The viscous and conductive heat fluxes are, respectively, 
\begin{subequations}
	\label{eq:heatfluxes}
	\begin{align}
	\tau_{ij} &= 2 \mu \left[S_{ij} + \frac{\secondcoefficient}{2 \mu} \frac{\partial u_k}{\partial x_k} \delta_{ij} \right]
	\label{subeq:hf1}
	\\
	q_j &= -\frac{\mu\,C_p}{\Pran} \frac{\partial}{\partial x_j} T
	\label{subeq:hf2}
	\end{align}
\end{subequations}
where $S_{ij}$ is the strain-rate tensor, given by $S_{ij}=(1/2) \left(\partial u_j/\partial x_i + \partial u_i /\partial x_j \right)$; $\Pran$ is the Prandtl number; and $\mu$ is the dynamic viscosity, given by $\mu = \mu_{\textrm{ref}}\left(T/T_\textrm{ref}\right)^{n_{\nu}}$, where $n_{\nu}$ is the viscosity power-law exponent and $\mu_{\textrm{ref}}$ is the reference viscosity. $\secondcoefficient$ is the second viscosity defined by \cref{e:secondviscosity}, $\mu_B \equiv \secondcoefficient + \frac{2}{3}\mu$, 
where $\mu_B$ is the effective bulk viscosity value capturing the combined effects of rotational and vibrational molecular relaxation.

Simulations have been carried out with the following gas properties: $\gamma=1.4$, $\rho_{\textrm{ref}} = 1.2\,\textrm{kg m}^{-3}$, $p_{\textrm{ref}} = 101\,325\,\textrm{Pa}$, $T_{\textrm{ref}}=300\,\textrm{K}$, $\mu_{\textrm{ref}}=1.98\times10^{-5} \, \textrm{kg}\, \textrm{m}^{-1} \textrm{s}^{-1}$,  $\Pran=0.72$, and $n_{\nu}=0.76$, valid for air.\cite{DeYiB_InternationalJournalHeatMassTransfer_1990} 

The governing equations are solved using \charlesx{}, a control-volume-based, finite-volume solver for the fully compressible Navier--Stokes equations on unstructured grids, developed as a joint-effort among researchers at Stanford University. \charlesx{} employs a three-stage, third-order Runge-Kutta time discretization and a grid-adaptive reconstruction strategy, blending a high-order polynomial interpolation with low-order upwind fluxes.\cite{HamMIM_2007} 
The code is parallelized using the Message Passing Interface (MPI) protocol and highly scalable on a large number of processors.\cite{BermejoMorenoBLBNJ_2013}

\subsection{\label{app:timedomain_lbm} Lattice Boltzmann Equations}

The LBM solver uses the Bhatnagar-Gross-Krook (BGK) collision model. 
Variables given in this section will be dimensionless in the LBM solver. 
The quadrature points selected for each lattice node are
\begin{align}
\vec{e}_i &= 
\begin{cases}
(0,0) & i=0\\ 
(1,0), (0, 1), (-1, 0), (0, -1) & i=1,2,3,4 \\
(1,1), (-1,1), (-1,-1), (1,-1) & i=5,6,7,8 \, ,
\end{cases} 
\end{align}
with corresponding weights of
\begin{align}
w_i &= 
\begin{cases}
4/9 & i=0\\ 
1/9 & i=1,2,3,4 \\
1/36 & i=5,6,7,8 \, , 
\end{cases} 
\end{align}
and a streaming and collision update model of 
\begin{align}
f_i &\left(\vec{x}^*+\vec{e}_i, t^*+1 \right) - f_i \left(\vec{x}^*, t^*\right) \nonumber \\&= -\frac{1}{\tau^*} \left[f_i \left(\vec{x}^*, t^*\right) -f_i^{eq} \left(\vec{x}^*, t^*\right)\right] \, , 
\end{align}
where $f_i$ is the particle distribution function for each streaming direction, $f_i^{eq}$ is the corresponding equilibrium distribution, and $\vec{x}^*$, $t^*$, and $\tau^*$ are the non-dimensional grid, time, and relaxation time, respectively. 
The lattice units are chosen to be $\Delta x^* = \Delta t^* = 1$, such that the non-dimensional lattice speed is $a_{0}^* = 1/\sqrt{3}$. 

The relaxation time $\tau^*$ determines the fluid kinematic viscosity, 
\begin{align}
\nu^* &= \left(\tau^*-1/2\right) \left(a^*_{0}\right)^2 = \frac{2\tau^*-1}{6}  \, .
\end{align}
For a channel of defined physical length
$L$ and $n_x$ grid points, in order to represent a simulation of physical kinematic viscosity $\nu_0$ and reference density $\rho_0$,
the conversion from lattice units to physical units is defined as
\begin{align}
x &= C_x x^* &
t &= C_t t^* \nonumber \\
\nu &= C_\nu \nu^* &
\rho &= C_{\rho} \rho^* \, ,
\end{align}
where 
\begin{align}
C_x &= \frac{L}{n_x}  & 
C_t &= \frac{C_x^2}{C_\nu} = \frac{C_x^2 \nu^*}{\nu_0} = \frac{2\tau^*-1}{6\nu_0} C_x^2 \nonumber
\\
C_\nu &= \frac{C_x^2}{C_t} & 
C_{\rho} &= \rho_0 \, .
\end{align}

\subsubsection{\label{sec:lbm:sc} LBM with MRT}

The lattice Boltzmann method in its simplest form assumes equal relaxation times, $\tau^*$, across multiple moments. 
Multiple relaxation time (MRT) models extend LBM by removing this constraint, allowing different and multiple relaxation times to be used instead. 
Typically, MRT models replace $\tau^*$ as used in the previous section with separate relaxation times for each value. 
Several of these relaxation times introduced by MRT correspond to   ``ghost modes,'' which have no basis in kinetic theory, but can enhance stability and indirectly affect fluid flow.\cite{AsinariK_PhysRevE_2009}
Nevertheless, some of these modes have direct physical connections to decoupling bulk and shear viscosity, and the presented model avoids tuning the MRT parameters unphysically. 
LBM with MRT and the proposed bulk viscosity model (\cref{sec:bv_model}) can fully capture acoustic attenuation in air.

For the chosen quadrature points and corresponding weights, the MRT model used is specified by the forward and inverse matrices, 
\begin{subequations}
\begin{align}
M = &\begin{pmatrix}
1 &  1 &  1 &  1 &  1 & 1 &  1 &  1 &  1 \\
-4 & -1 & -1 & -1 & -1 & 2 &  2 &  2 &  2 \\
4 & -2 & -2 & -2 & -2 & 1 &  1 &  1 &  1 \\
0 &  1 &  0 & -1 &  0 & 1 & -1 & -1 &  1 \\
0 & -2 &  0 &  2 &  0 & 1 & -1 & -1 &  1 \\
0 &  0 &  1 &  0 & -1 & 1 &  1 & -1 & -1 \\
0 &  0 & -2 &  0 &  2 & 1 &  1 & -1 & -1 \\
0 &  1 & -1 &  1 & -1 & 0 &  0 &  0 &  0 \\
0 &  0 &  0 &  0 &  0 & 1 & -1 &  1 & -1 
\end{pmatrix} 
\displaybreak[0]
\\ 
M^{-1} = \frac{1}{36} &\begin{pmatrix}
4 & -4 &  4 &  0 &  0 &  0 &  0 &  0 &  0 \\
4 & -1 & -2 &  6 & -6 &  0 &  0 &  9 &  0 \\
4 & -1 & -2 &  0 &  0 &  6 & -6 & -9 &  0 \\
4 & -1 & -2 & -6 &  6 &  0 &  0 &  9 &  0 \\
4 & -1 & -2 &  0 &  0 & -6 &  6 & -9 &  0 \\
4 &  2 &  1 &  6 &  3 &  6 &  3 &  0 &  9 \\
4 &  2 &  1 & -6 & -3 &  6 &  3 &  0 & -9 \\
4 &  2 &  1 & -6 & -3 & -6 & -3 &  0 &  9 \\
4 &  2 &  1 &  6 &  3 & -6 & -3 &  0 & -9 
\end{pmatrix} 
\end{align}
\end{subequations}
and the diagonalization of
\begin{align}
S &= \textrm{diag} \left( 0, s_2, 1.4, 0, s_5, 0, s_7, s_8, s_9 \right)\, ,
\end{align}
where it is noted that $s_1, s_4, s_6$ can be set, without loss of generality, to 0, as mass and momentum are necessarily conserved, $s_5=s_7$ is required and chosen to be 1.2, and $s_8=s_9$ is also required and chosen to be $1/\tau^*$. 

The bulk viscosity is a tuned parameter and is related to $s_2$; in the MRT model, the shear and bulk kinematic viscosities are, respectively, 
\begin{align}
\nu^* &= \frac{\frac{2}{s_8}-1}{6} =\frac{2\tau^*-1}{6}  \\
\nu^*_{B} &= \frac{\frac{2}{s_2}-1}{6} \, . 
\end{align}
For implementation, the ratio of bulk viscosity to shear (dynamic) viscosity, $\mu_B/\mu$, is used to select $s_2$.

\end{document}